\documentclass{emulateapj}

\shorttitle{High-Redshift Galaxies in Hydro Simulations}
\shortauthors{Finlator et al.}

\begin{document}
\title{The Physical and Photometric Properties of High-Redshift Galaxies in Cosmological Hydrodynamic Simulations}
\author{K. Finlator\altaffilmark{1}\email{kfinlator@as.arizona.edu}}
\author{R. Dav{\' e}\altaffilmark{1}, C. Papovich\altaffilmark{1}}
\author{L. Hernquist\altaffilmark{2}}
\altaffiltext{1}{University of Arizona, Department of Astronomy, Tuscon, AZ 85721}
\altaffiltext{2}{Harvard-Smithsonian Center for Astrophysics, 60 Garden Street, Cambridge, MA 02138}

%Commands
\newcommand{\mstar}{$M_{\star}$ }
\newcommand{\ud}{\mbox{\ d}}
\newcommand{\msun}{\mbox{M}_{\sun}}
\newcommand{\zsun}{\mbox{Z}_{\sun}}
\newcommand{\lgmstar}{\log(M_*/\msun)}  
\newcommand{\lcdm}{$\Lambda$CDM}
\newcommand{\hinv}{h^{-1}}
\newcommand{\fixme}[1]{\textbf{\emph{ FIXME: #1}}}
\newcommand{\parname}[1]{\noindent\emph{\underline{#1:} }}
\newcommand{\rd} {$E(B-V)$ distribution}
\newcommand{\lf} {luminosity function}
\newcommand{\ld} {luminosity density}
\newcommand{\lnu} {\rho_{\nu}}
\newcommand{\sfr} {star formation rate}
\newcommand{\lbg} {Lyman Break Galaxy}
\newcommand{\ebv} {$E(B-V)$}
\newcommand{\smyr} {\msun \mbox{ yr}^{-1}} % in math mode
\begin{abstract}
We study the physical and photometric properties of galaxies at
$z=4$ in cosmological hydrodynamic simulations of a \lcdm~universe.
We focus on galaxies satisfying the ``B-dropout" criteria of the Great
Observatories Origins Survey (GOODS).  Our goals are: (1) to study the
nature of high-redshift galaxies; (2) to test the simulations against
published measurements of high-redshift galaxies; (3) to find relations
between photometric measurements by {\it HST}/ACS (0.4 -- 1 $\mu$m) 
and {\it Spitzer}/IRAC (3.6 -- 8 $\mu$m)
and the intrinsic physical properties of GOODS ``B-dropouts" such as
stellar mass, stellar age, dust reddening, and star-formation rate; and
(4) to assess how representative the GOODS survey is at this epoch.
Our simulations predict that high-redshift galaxies show strong
correlations in star formation rate versus stellar mass, and weaker
correlations versus environment and age, such that GOODS galaxies are
predicted to be the most massive, most rapidly star-forming galaxies
at that epoch, living preferentially in dense regions.  The simulated
rest-frame UV luminosity function (LF) and integrated luminosity density
are in broad agreement with observations at $z\sim 4$.  The predicted
faint end slope is intrinsically steep, but becomes shallower and is in
reasonable agreement with data once GOODS selection criteria are imposed.
The predicted rest-frame optical (observed $3.6\mu$) LF is similar to
the rest-frame UV function, shifted roughly one magnitude (AB) brighter.
We predict that GOODS detects less than 50\% of the total stellar mass density
formed in galaxies more massive than $\log(M_*/M_{\odot}) > 8.7$ by $z=4$,
mainly because of brightness limits in the {\it HST}/ACS bands.  Most of
these results are somewhat sensitive to the prescription used to model
the effects of dust extinction.  We develop a physically-motivated model
that is based on taking the simulation-predicted metallicities and using
the dust-metallicity relation calibrated locally from SDSS.  This model
generally produces results in good agreement with observations, although
it produces a modest excess of bright, rapidly star-forming galaxies.
The slope of the predicted stellar mass-metallicity relation is in excellent 
agreement with low-redshift measurements of the stellar mass--gas-phase 
metallicity relation such that over two decades in stellar mass, galaxies 
are less enriched than low-redshift galaxies of similar stellar mass by 
roughly 0.6 dex.  The most rapidly
star forming galaxies in our simulations have rates exceeding 1000~$\smyr$,
similar to observed sub-mm galaxies.  These galaxies are not starbursts,
however, as their star formation rates show at most a mild excess 
($\sim 2-3\times$) over the star formation rates that would be expected for 
their stellar mass.  It is possible that the observable counterparts
to these bright galaxies do not follow our dust prescription and are instead 
heavily extinguished.  The overall distribution of dust reddening and mean 
stellar age may be constrained from color-color plots although the specific 
value for each galaxy cannot.
\end{abstract}

\section{INTRODUCTION} \label{intro}

According to the currently-favored hierarchical model of structure formation,
the first generations of stars began forming in low-mass galaxies at
very high redshift.  These galaxies then grew by accreting gas from their
surrounding intergalactic medium (IGM), and merged into larger galaxies that we observe today.
The details of how this process operates are still not well understood.
Numerical simulations that include a range of physical processes
believed to govern galaxy formation allow the hierarchical scenario
to be tested through detailed comparisons with observed galaxies.
This yields insights into the nature of galaxy formation and assists
with interpreting observations of galaxies~\citep{Weinberg2002}.

The current generation of deep surveys from the optical to the radio
are giving us an unprecedented view of the high-redshift ($z\ga 3$)
universe, and as a result are providing new and stringent tests on models
of galaxy formation.  Recent years have also seen a rapid growth in the
power and sophistication of numerical simulations of galaxy formation,
so that detailed comparisons are ever more meaningful and illustrative.
For these reasons, the time is ripe for a detailed investigation into
the physical nature and observable properties of high-redshift galaxies
in cosmological simulations of galaxy formation.

The most effective method for finding high-redshift
objects for observational study is the Lyman break selection
technique~\citep{SH1993,Steidel2}, exploiting the large differential in
flux across the rest-frame Lyman limit owing to internal absorption
within the galaxy as well as strong absorption by intergalactic
neutral hydrogen along the line of sight.  Together with the fact
that high-redshift galaxies are necessarily young and therefore blue,
this endows high-redshift galaxies with distinctive colors that make
photometric selection straightforward.  Spectroscopic followup has shown
that this technique is highly effective at isolating Lyman break galaxies
(LBGs) while suffering little contamination from lower redshifts.

The first attempts to simulate LBGs used semi-analytic
mod\-els~\citep[SAMs; see e.g.][]{Kauf1,SPF1}, which employ merger trees
or gravity-only numerical simulations to follow the growth of dark
matter perturbations, and use analytic prescriptions to
account for the remaining physical processes involved in forming galaxies
such as merging, cooling, and feedback.  These prescriptions give SAMs
flexibility and low computational cost, but their predictions of galaxy
properties consequently rely on a variety of simplifying assumptions.
Hydrodynamic simulations complement the SAM approach by allowing galaxies
to be simulated with minimal assumptions regarding the dynamics of gas and
dark matter, but are computationally costly and are limited by numerical
resolution and volume effects.

Different SAM prescriptions have yielded different physical models
for LBGs, with some SAMs~\citep[notably][]{SPF1} indicating that
a significant fraction of LBGs are merger-induced starbursts,
while others~\citep{Baugh1998} holding that LBGs are the most
massive galaxies to have formed at these epochs.  Hydrodynamic
simulations, meanwhile, have consistently pointed towards
the latter scenario~\citep{Dave99,N1,N2005a,Weinberg2002}.  To date,
efforts to infer the physical properties of LBGs via population
synthesis modeling have tended to support the massive galaxy
hypothesis~\citep{Shapley2005,Shapley1,Papovich2,Barmby1} although only a small
subset of LBGs have had their physical properties studied in detail.
In this paper, we elaborate on the predictions of the massive galaxy
model in order to facilitate further comparison with current and future
observational work.

While determining the basic properties of LBGs as a constraint on
galaxy evolution models is an important goal unto itself, it is also a
critical step towards constraining specific volume-averaged quantities
such as galaxy number densities, luminosity functions (LFs), luminosity
density ($\lnu$), and the history of cosmic star formation~\citep{Madau2,lil97}.
All models of galaxy formation predict a substantial population of 
sources at high redshift that cannot
be detected via the Lyman break technique, although they differ as to
this population's properties.  For example, the SAM of~\citet[Figure
2]{Idzi1} predicts a correlation between rest-frame UV flux and stellar
mass that is substantially weaker than the trend from the hydrodynamic
model of~\citet[Figure 3; see also our Figure~\ref{phys_phot}]{Dave99}.
Thus, the SAM used by~\citet{Idzi1} predicts that UV-faint and UV-bright
galaxies are more similar in stellar mass---implying that a larger
fraction of total formed stellar mass may be ``missed"---than would be
predicted by the hydrodynamic model of~\citet{Dave99}.  Testing the
observable properties of LBGs against models will help in accurately
assessing their contribution to global stellar mass and star formation
rate budget, as well as their relationship to various other galaxy
populations at those epochs.

Currently the single greatest difficulty in modeling LBGs accurately
is the poorly-understood impact of extinction owing to dust.  Nearly all
star-forming galaxies at high redshift are seen to suffer from some
level of extinction~\citep{AS1,Shapley1}, usually expressed as a
color excess \ebv.  To date, constraints on reddening at high redshift
have been obtained by fitting population synthesis models to observed
LBGs~\citep{Shapley1,Papovich2}.  Unfortunately, this means
the resulting distributions of \ebv~are sensitive to selection biases
caused by the reddening itself, making it difficult to infer how many
galaxies may be missed by such studies.  In particular, as LBG selection
criteria are biased against heavily reddened objects with $E(B-V) >
0.3$~\citep{daddi2004}, LBG studies are likely to underpredict the median
\ebv~suffered by the complete galaxy population.  In support of this,
recent observations suggest that a substantial fraction of the total star
formation occurring at high redshift occurs in highly reddened galaxies
that are too faint in rest-frame UV to be selected as LBGs but can be
selected either via their rest-frame optical colors~\citep{daddi2004}
or in the sub-millimeter~\citep{Chapman1,Chapman2}.  The effects of dust on
the most massive galaxies ($\lgmstar > 11$) may not even follow simple
attenuation laws such as that of~\citet{Calz1} (\citealt{Chapman2}; Papovich
et al.\ 2005, in preparation).  Insight into the
effects of dust on a galaxy's observed spectral energy distribution
(SED) and the relationship of extinction to its more fundamental
properties such as stellar mass, metallicity, and star formation rate
is necessary for determining the completeness of the current observational
census of high-redshift galaxies.  In this paper we study a range of
reddening models, comparing to observations where possible, and provide
an observationally-motivated reddening prescription that is the most
sophisticated one used to-date in simulations.

In order to perform a meaningful statistical comparison to understand
these issues, a large sample of observed LBGs is required.  The Great
Observatories Origins Deep Survey (GOODS) has provided an unprecedented
view into the high-redshift universe~\citep{Giavalisco1}.  GOODS probes to
a depth ($i_{AB}<27.5$) comparable to previous space-based observations
such as the Hubble Deep Field~\citep{Williams1996} while covering an
area ($320$~arcmin$^2$) comparable to shallower ground-based studies.
Combining {\it HST}/ACS data and {\it Spitzer}/IRAC data allows
constraints to be placed on the star formation rate, stellar mass,
extinction, and other properties of galaxies out to $z\sim 4$ and
beyond~\citep{Giavalisco2,Papovich1}.  The field size is even large enough to 
study clustering~\citep{Roche1}, and X-ray observations have provided auxiliary
data on AGN activity and star formation~\citep{Lehmer1}.  At present,
1115 LBGs at $z\sim4$ have been isolated from GOODS data~\citep{Giavalisco2}.  
Our simulations have sufficient dynamic range to numerically 
resolve the entire ``B-dropout" sample within a comoving 
cosmological volume that is comparable to the volume probed by
GOODS at $z\sim4$.  In addition, these simulations have been shown to
broadly match the observed cosmic star formation rate density~\citep{SH1,HS},
and the rest-frame UV luminosity functions of LBGs when a moderate amount
of dust extinction is assumed~\citep{N1,N1erratum,N2005a,Night1}.

In this work we present comparisons between the ensemble statistical
properties of high-redshift galaxies in our simulations and in the
GOODS data set; examine the basic physical properties of a simulated
GOODS sample and discuss the trends that emerge; investigate how the
photometric properties of the GOODS sample relate to the underlying
physical properties; determine how the model parameters affect the
observed color-magnitude and color-color relations; discuss the extent
to which the observed GOODS sample will be representative of the complete
galaxy population at $z=4$; and compare our results with those of previous
simulations and semi-analytic models.  Our work provides a thorough
investigation into using the observed photometric properties of
high-redshift
galaxies as a new frontier for testing galaxy formation models.

The paper is organized as follows: In Section~\ref{simdescript}
we summarize the input physics in the simulation, our procedure for
identifying galaxies and computing their observed photometry, and
our procedure for adding dust reddening to the simulated galaxies.
In Section~\ref{resolution} we discuss several tests that we used to
verify that our simulated galaxies' properties are numerically resolved.
In Section~\ref{basicprops} we summarize the ensemble statistical
properties of the simulated sample including incompleteness owing to the
GOODS $z\sim4$ selection function, the integrated luminosity density,
and LFs.  In Section~\ref{sec:detailedprops} we present a number of plots 
comparing physical and photometric properties and briefly introduce the 
most rapidly star forming galaxies in the simulated sample.  Finally, in
Section~\ref{sec:conclusion} we present our conclusions.

\section{SIMULATIONS AND SAMPLE DEFINITION} \label{simdescript}
\subsection{Simulations} \label{subsec:simulations}

Our work is based on cosmological hydrodynamics
simulations done using \textit{GADGET-2}~\citep{S05}.
\textit{GADGET-2}\footnote{\raggedright Publically available at \url{http://www.mpa-garching.mpg.de/gadget/}}
models the evolution of a system of particles under gravity using
a Tree-particle-mesh solver and hydrodynamical forces using an
entropy-conservative formulation \citep{SH02}
of smoothed particle hydrodynamics
(SPH).  Our particular simulation is the G6 simulation, an extension
of the G-series runs described in~\citet{SH1}.  It was started at $z =
79$ with $486^3$ dark matter and $486^3$ SPH particles in a cube with
periodic boundary conditions and comoving side length $100 \hinv$ Mpc.
The simulation has a SPH particle mass resolution of $9.7 \times 10^7 \; \hinv
\; \msun$ and an equivalent plummer softening length of $5.33 \; \hinv$
kpc.  For comparison, we also employ the D5 simulation of~\citet{SH1},
which has higher resolution than G6, but in a smaller volume (see below).
They were run using a concordant \lcdm~cosmology~\citep{Spergel2003}:
($\Omega_{\mbox{m}}$, $\Omega_{\Lambda}$, $\Omega_{\mbox{b}}$, $\sigma_8$,
$h$) = (0.3, 0.7, 0.04, 0.9, 0.7); we assume this cosmology throughout.
These simulations were run on the Athlon-MP cluster at the Center
for Parallel Astrophysical Computing (CPAC) at the Harvard-Smithsonian
Center for Astrophysics.

Galaxy star formation histories are the key output of \textit{GADGET-2} that
we use to model the observable properties of high redshift galaxies.
\textit{GADGET-2} forms stars using a subgrid multiphase model for
the interstellar medium~\citep[ISM;][]{SH2}.  If a SPH particle's
density surpasses a threshold value based on the local thermal Jeans
mass, it acquires a two-phase ISM consisting of a hot ambient medium in
pressure equilibrium with a cold dense medium~\citep[based on][]{MO1977}.
A thermal instability causes gas in the hot phase to cool and condense
into the cold medium.  The cold medium, in turn, forms stars on
a timescale that is calibrated to match the observed star formation
rate in local spirals~\citep{Kennicutt1998}.  A fraction of these stars
forms supernovae that enrich and deposit thermal energy into the
hot phase of the ISM.  
The thermal feedback causes cold gas to evaporate back into
the ambient hot medium, depleting the gas reservoir from which stars form.
In this way, \textit{GADGET-2} computes a subresolution, self-regulated
star formation and feedback cycle within each star-forming SPH particle.
The code also tracks the metal enrichment of each SPH particle as the
simulation evolves. Star particles are spawned in two steps from each
SPH particle by~\textit{GADGET-2}, so that each star particle has half the
mass of the SPH particles.  A star particle inherits the metallicity of
its parent SPH particle, and does not change thereafter.

Unfortunately, by itself this multiphase ISM model cannot
prevent the simulation from overproducing the observed stellar
mass density in the universe, so some form of kinetic feedback is
needed~\citep{Balogh2001,K1,SH2,HS}.  For this reason, the simulation also
includes a phenomenological Monte Carlo prescription for generating
galactic superwinds as observed in local starburst galaxies.  The code
selects SPH particles with a star-forming ISM and gives a ``kick" to their
velocity vectors in the direction of the cross product of the particle's
velocity and acceleration, creating broadly bipolar outflows.  The model
assumes that the mass outflow rate from star forming regions is twice the
local star formation rate, and the velocity of each ``kick" is set to 484
km s$^{-1}$ in order to reproduce the observed space density of stellar
mass at low redshift~\citep{SH1}.  Besides suppressing star formation,
this model creates qualitatively realistic large-scale galactic outflows
that chemically enrich the IGM as proposed by~\citet{Aguirre2001a,Aguirre2001b},
although~\citet{Aguirre2005} suggest that this superwind model may require
some tuning.

\subsection{Galaxy Identification} \label{subsec:groupfinder}

We use the Spline Kernel Interpolative DENMAX 
(SKID)\footnote{\raggedright \url{http://www-hpcc.astro.washington.edu/tools/skid.html}}
group finder to identify gravitationally bound groups of star and gas
particles as galaxies.  SKID operates in several steps, which may be
summarized as follows: (1) slide particles along the gradient of the
initial baryonic density field until they are confined into subgroups
about the initial density peaks; (2) combine subgroups into groups using
a friends-of-friends algorithm; (3) reject particles from each group
that are not energetically bound to that group; and (4) reject groups
that have fewer than a minimum number of members.  We used a 16-member
cutoff when running SKID but applied a much more conservative cut at
64 star particles later on (\S~\ref{resolution}).  We employ
SKID rather than another group finder because test runs suggest that it
separates satellite galaxies from main galaxies more accurately than halo
finders that select objects based on dark matter alone.  As a check we
compared our results to those of~\citet{N1}, who used a different group
finder with the same simulations, and found no significant differences
(\S~\ref{resolution}).

\subsection{Simulated Galaxy SEDs} \label{subsec:bc03}

For each SKID-identified galaxy, we treat each star particle as a
single stellar population (SSP) formed at the time the star particle was
spawned.  We then determine the broadband photometric properties of
each galaxy by processing its star formation history through the GALAXEV
library of evolutionary population synthesis models~\citep{BC1}, 
redshifting the resulting spectrum, and
convolving it with our photometric filters.  We use
a~\citet{Chabrier2003} IMF.  The filters we use here are the ACS F435W,
F606W, F775W, and F850LP (hereafter, $B_{435}$, $V_{606}$, $i_{775}$,
and $z_{850}$, respectively); the \textit{VLT}/ISAAC $J$, $H$, and $K_s$;
and the IRAC 3.6$\mu$, 4.5$\mu$, 5.8$\mu$, and 8.0$\mu$ ([3.6], [4.5],
[5.8], and [8.0], respectively) bands.  Throughout this work we use AB
magnitudes~\citep{okegunn1983}.

As the GALAXEV library covers a range of metallicities, we must decide
which one to use.  Figure~\ref{metallicity_reddening} shows that simulated
galaxies' metallicities $\log(Z/\zsun)$ at $z=4$ fall between $-0.7$
and $-0.4$.  Tests showed that using the $\log(Z/\zsun) = -0.7$ models for
all galaxies produces errors of less than 0.1 magnitudes for all filters
of interest, which is good enough to preserve any important qualitative
trends in the data.  Therefore, we present results only from models with
$\log(Z/\zsun) = -0.7$.  Since much of the population synthesis work in
the literature uses models with solar metallicity, we note that using
the solar metallicity models would redden our galaxies by less than 0.01
magnitudes in $K_s$ - [3.6] ($\approx$~rest-frame $B-V$) and dim them
by 0.2 and 0.3 magnitudes in rest-frame optical (observed [3.6]) and 
UV (observed $i_{775}$), respectively.

\subsection{Reddening from Dust and the IGM} \label{subsec:reddening}

\begin{figure*}[ht]
\epsscale{1.00}
\plotone{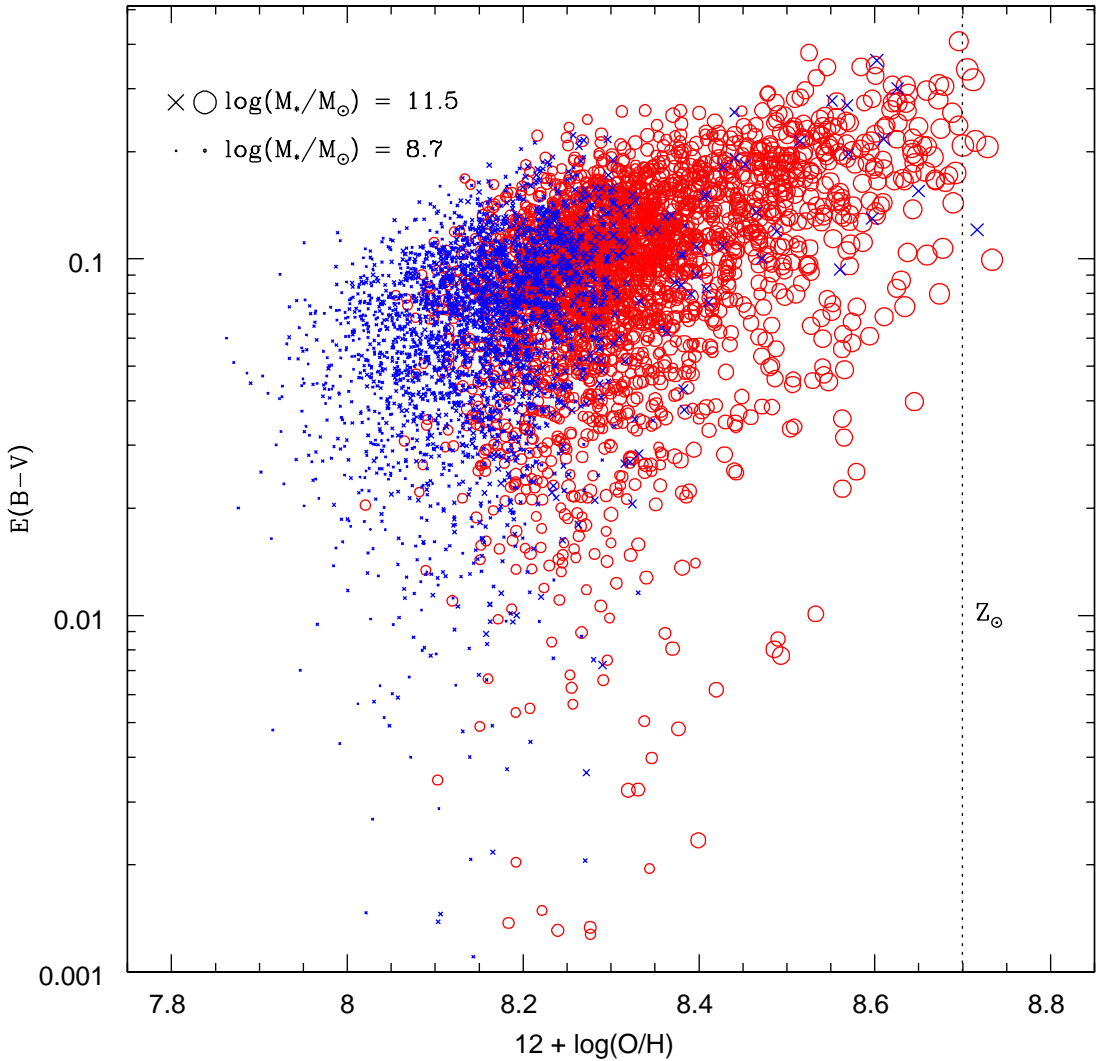}
\caption[]{The applied $E(B-V)$ distribution.  The x-axis gives the metallicity of each
galaxy while the y-axis gives the applied dust reddening using the~\citet{Calz1}
reddening curve.  Solar metallicity is indicated by the dotted line.  Red circles and 
blue crosses are from the G6 and D5 simulations, respectively.  Point size scales linearly 
with $\lgmstar$ as indicated.  Note that more massive objects suffer significantly 
more dust reddening in this prescription.}
\label{metallicity_reddening}
\end{figure*}

The most common method for inferring the star formation rates of galaxies
involves measuring the rest-frame UV light emitted by short-lived massive
stars.  Unfortunately, evidence suggests that much of the rest-frame
UV light from star-forming galaxies is absorbed by a dusty interstellar
medium and re-emitted at far-infrared wavelengths~\citep{AS1}.  Both the
amount of dust in each galaxy and the variation of the extinction as
a function of wavelength (the ``$E(B-V)$ distribution" and ``reddening
law," respectively) are poorly constrained for LBGs at present.  Available
observational constraints tend to assume a~\citet{Calz1} reddening curve,
which is parametrized in terms of the color excess \ebv.  Several studies
have derived $E(B-V)$ distributions using this law from population
synthesis~\citep{Shapley1} or UV spectral slope~\citep{AS1,Ouchi1}
techniques.  Consensus has centered on a median value of \ebv$\sim 0.15$ 
with some scatter for LBGs at $z\sim3$, corresponding to a median factor 
of 3--5 extinction in the UV.  For consistency with past work, we will 
employ the ~\citet{Calz1} reddening law, although we briefly consider 
the~\citet{CF2000} law as well.

Given the reddening law, a color excess \ebv\ must be chosen
for each galaxy.  As argued above, it is not appropriate to simply apply
an observed $E(B-V)$ distribution such as that of~\citet{Shapley1} to our
sample, as there is no \emph{a priori} reason to expect that the intrinsic
and observed $E(B-V)$ distributions are the same.  The intrinsic $E(B-V)$
distribution could, for example, include a significant population of
highly reddened objects; the dust in such objects would render their
observable rest-frame UV colors too red for them to pass the LBG color
cuts that have been used to date~\citep[see e.g. the $BzK$ samples
of][]{daddi2004}.

The state-of-the-art in intrinsic $E(B-V)$ distributions is more art
than science.  \citet{N1} do not assume an intrinsic $E(B-V)$ distribution,
but rather plot their results for several different uniform values of
\ebv~and note that the data match the simulated LFs and number counts well
if the typical \ebv~at redshifts $z \sim$ 3--5 is 0.15.  By contrast,
recent SAMs~\citep{SPF1,Idzi1} scale reddening with UV luminosity in a
way that mimics the correlations observed in nearby starburst galaxies;
we will explore this \rd, but we note that it is not well-motivated for
our sample because most of our simulated LBGs are not starbursts.

Here, we employ the novel approach of inferring each galaxy's dust
content \ebv~from its metallicity.  It is natural to expect a trend
between reddening and metallicity since dust grains are made of metals;
indeed, such a trend was first observed in IUE spectra of starbursts
by~\citet{SB1994}.  We use a correlation observed between the reddening
and metallicity in the Sloan Digital Sky Survey (SDSS) main galaxy sample
(C. Tremonti, private communication), extrapolating the mean trend
to each galaxy's metallicity.  For each galaxy, we also add a Gaussian
scatter $\delta E$ with variance equal to one half of the mean color
excess $\langle E(B-V)\rangle$ at that galaxy's metallicity; this 
produces a good match to scatter observed in the SDSS sample.  
The final relation is
\begin{eqnarray} \label{eqn:ebv}
E(B-V) = 9.0 \times Z^{0.9} + \delta E,
\end{eqnarray}
where $Z$ is the galaxy's mean stellar metallicity (expressed as a ratio
of the mass in metals to the total stellar mass) and the 
scatter $\delta E$ is of a Gaussian form given by
\begin{eqnarray}
f(\delta E) \ud (\delta E) & = & \frac{1}{\frac{1}{2}\langle E(B-V)\rangle\sqrt{2\pi}} \nonumber \\
& \times & \exp\left[-\frac{1}{2} \left(\frac{E(B-V)}{\frac{1}{2}\langle E(B-V)\rangle}\right)^2\right].
\end{eqnarray}

Figure~\ref{metallicity_reddening} shows the results of this prescription
applied to all simulated galaxies by giving the average metallicity
of the star particles in each simulated galaxy versus the applied
color excess \ebv.  The solar abundance is marked with a dotted line.
While the simulation does not trace the abundances of individual metal
species, we report the metallicity as an oxygen abundance since this is
a common tracer of other metals.

\begin{figure}[ht]
\epsscale{1.00}
\plotone{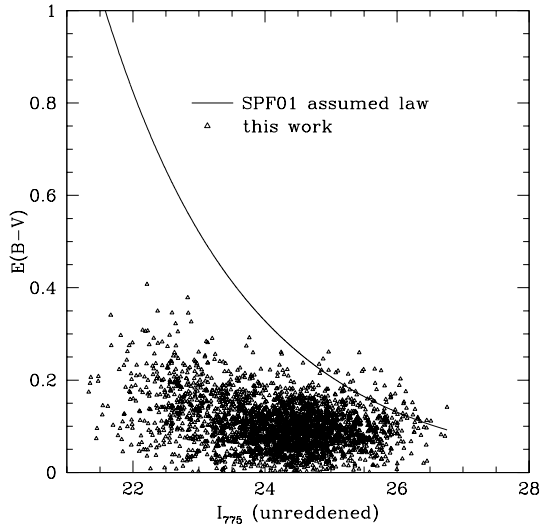}
\caption[]{Applied reddening \ebv~versus unextinguished flux in $i_{775}$ 
(rest-frame 1500 \AA) for all resolved galaxies. Open triangles give our 
fiducial prescription and the solid line gives the mean of the prescription 
used in~\citet{SPF1}.  Our applied reddening scales much less strongly 
with star formation rate than the prescription used in~\citet{SPF1}.}
\label{ebv-fiducial-spf01}
\end{figure}

As we will show, this method yields results that are in broad agreement
with other methods, such as applying a uniform \ebv~to all galaxies
or picking \ebv~from a Gaussian distribution.  It naturally produces a
weak correlation between mass and reddening since galaxies with higher
stellar mass tend to have higher metallicities.  In principle, it also
has the potential to produce massive, evolved galaxies with no gas and
(unphysically) significant dust.  However, inspection of our simulated
B-dropout sample revealed no massive galaxies with abnormally low gas
surface density or star formation rate.  The lack of massive gas-free
galaxies at low redshifts would certainly indicate a failing of the
simulation, but it is unclear whether it is a failing at $z\sim4$.
In any case, it does remove the necessity of further refining our
reddening prescription at this point.

Figure~\ref{ebv-fiducial-spf01} shows our fiducial \rd~and compares it
with the mean of the distribution used in~\citet{SPF1} and~\citet{Idzi1}.
The abscissa gives the observed flux in $i_{775}$ before reddening
is applied and the ordinate gives the applied reddening \ebv.
To generate the curve, we adapted the~\citet{SPF1} prescription
assuming a $\lambda^{-0.7}$ reddening curve and $R_V = A_V/E(B-V) =
4.05$~\citep{Calz1} to derive $E(B-V) = 0.189 \times 10^{0.2(24.98 -
i_{775})}$.  Using the mean relation between the unextinguished flux in
$i_{775}$ and the intrinsic star formation rate $\dot{M_*}$ for our
galaxies that we derive as described in \S~\ref{subsec:bc03},
this can also be written as $E(B-V) = 0.189 \times 10^{(-0.44 +
0.52\log(\dot{M_*}))}$ where $\dot{M_*}$ is in $\smyr$.  While our
prescription naturally leads to UV-brighter objects having greater
extinction, our mean trend is not nearly as strong as theirs.

For comparison, we explore several other \rd{s}.  In one, each galaxy's
reddening is determined randomly via a Gaussian distribution of \ebv~with
mean and standard deviation 0.15 but truncated at 0.0 so that no galaxy 
suffers negative extinction\footnote{Note, however, that some 
authors~\citep[e.g.,][]{Steidel1} have derived apparently negative 
values of \ebv~owing to significant Lyman alpha emission boosting the 
rest-frame UV flux.}; this is referred to hereafter as the ``random" 
\rd~and is similar to the distribution used by~\citet{Night1}.  We also 
use a sample in which each galaxy's reddening is set to 0.12 (the ``flat"
\rd); this number was chosen to produce reasonable agreement between
the simulated and observed luminosity densities in the rest-frame UV
(cf.\ \S~\ref{sec:lumDensity}).  Unless otherwise noted, we employ
our fiducial \rd~for all of our results.

Finally, in addition to dust extinction we must model attenuation owing to
the IGM along the line of sight to the galaxy.  We use the prescription
for the mean attenuation given by~\citet{Madau1}, which accounts for a
stochastic distribution of optically thin Lyman-$\alpha$ forest clouds
as well as optically thick Lyman limit systems.  For our $z=4$ galaxies,
this prescription suppresses the simulated fluxes in the observed
$B_{435}$ and $V_{606}$ bands by roughly $1.6\pm0.1$ and $0.4\pm0.1$ 
magnitudes, respectively (the scatter owes to the variation in the shapes 
of the intrinsic galaxy SEDS), and does not affect redder bands at all.

\section{NUMERICAL RESOLUTION EFFECTS} \label{resolution}

There are a number of ways in which cosmological N-body simulations
can fail to resolve the physical properties of galaxies.  As a relevant
example, Figure 10 in~\citet{SH1} shows how the simulated cosmic star
formation rate density at a given redshift varies with mass resolution.
Broadly, using higher mass resolution allows lower-mass objects to be
resolved, and allows structure formation at all scales to be resolved
earlier in a simulation; by contrast, simulating a larger cosmological
volume typically entails using lower mass resolution but allows rare,
massive objects to be simulated.  For these reasons, the cosmic star
formation rate density in the G6 simulation lies below the correct value
inferred from higher resolution simulations until slightly after $z=4$.
During this time, some of its galaxies may possess lower stellar mass and
bluer colors than the converged values.  Since we are mainly considering
the properties of the largest galaxies that collapsed earliest, we expect
our star formation rates to show better convergence than the overall
galaxy population.  However, as a check that our simulated sample's
properties are resolved, we repeat our analysis with the higher-resolution
D5 simulation of~\citet{SH1}, whose cosmic star formation rate density
converges to the resolved value before $z=4$.  D5 was started at $z = 159$
with $324^3$ dark matter and $324^3$ SPH particles in a cube of comoving
side length  $33.75 \hinv$Mpc.  It has a SPH particle mass resolution
of $1.26 \times 10^7 \; \hinv \msun$ and an equivalent Plummer softening
length of of $4.17 \hinv$kpc, and uses the same cosmology as G6.

\begin{figure*}[ht]
\epsscale{1.00}
\plotone{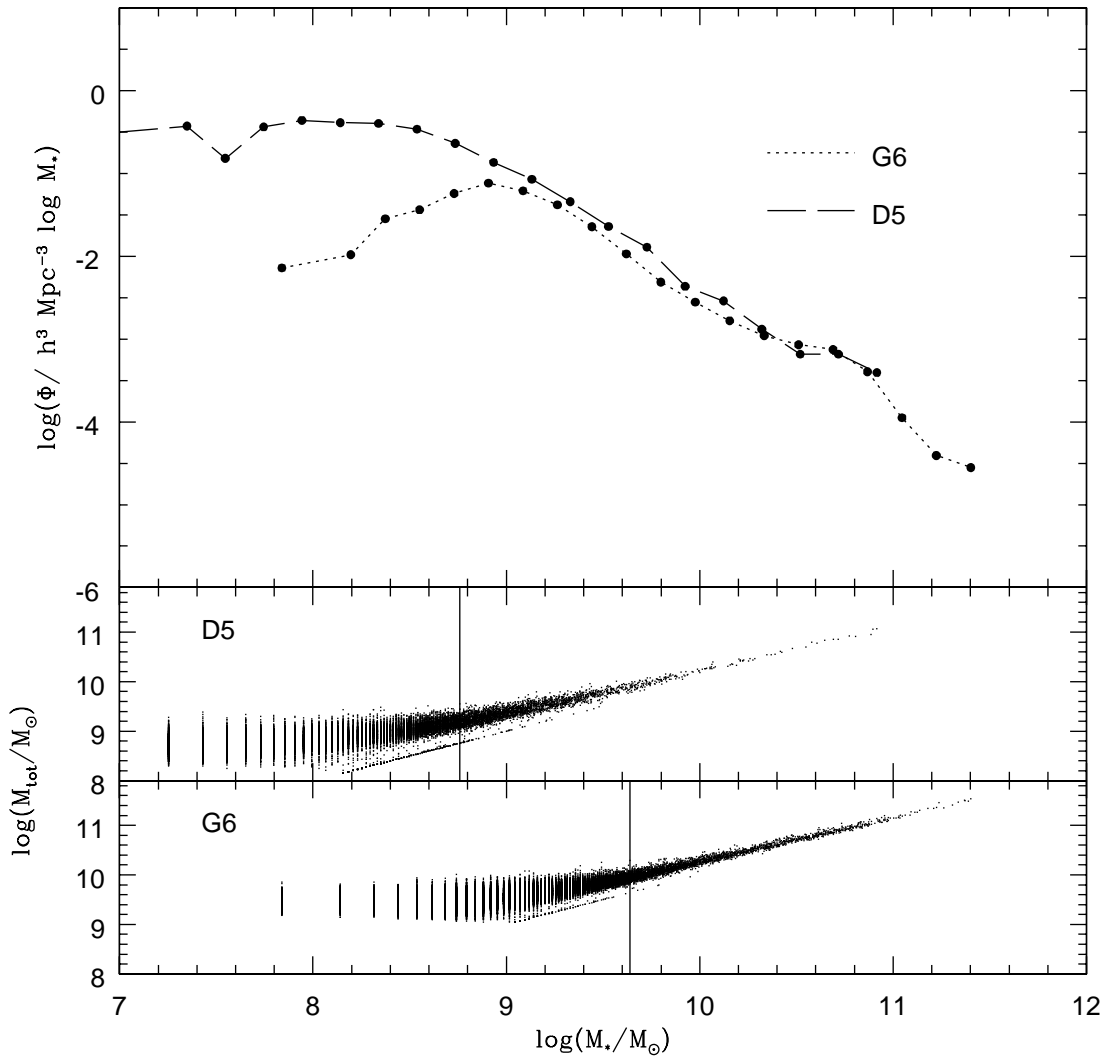}
\caption[]{(top) The mass functions of all galaxies found by SKID in the G6 and
D5 simulations. (middle and bottom) Total baryonic mass versus stellar mass
for all simulated galaxies in the D5 (middle) and G6 (bottom) simulations.  
The discretized behavior at low stellar masses results from finite mass 
resolution.  The heavy vertical lines indicate where our 64 star particle 
mass resolution cut falls in each simulation.  Our stellar mass cut removes
galaxies whose star formation histories are not well-resolved owing to the
stochasticity in the star formation prescription.}
\label{mass_function}
\end{figure*}

\parname{Mass Resolution} We determined the simulated mass resolution
by comparing the stellar mass functions for the G6 and D5 simulations
(Figure~\ref{mass_function}, top panel).  It is tempting to assume that
each simulation resolves the properties of all galaxies whose stellar
mass exceeds the mass at which the simulated stellar mass function tracks
that of the higher resolution simulation; Figure~\ref{mass_function}
in this case tells us that the G6 resolves galaxies with stellar mass
above $\lgmstar = 8.5$.  Unfortunately, since we are interested in
star formation histories of galaxies, this criterion is insufficient.
The bottom panels of Figure~\ref{mass_function} plot the total baryonic
mass versus stellar mass for the two simulations.  Inspection shows 
that the correlation between total baryonic and stellar mass is 
tight at high stellar masses but becomes noisy at lower 
stellar masses, well before the stellar mass function turns 
over.  This stochasticity, arising as a result of the stochastic
prescription for star formation in the code, would result in unphysical
scatter in our simulated star formation histories.  To avoid this, we
impose a stricter minimum stellar mass cut for G6 at 64 star particles or
$\lgmstar = 9.64$ (the solid line in the plot), which is slightly below
the stellar mass observed for typical ($L^*$) LBGs at 
$z\sim3$~\citep{Shapley1,Papovich2}.  Above this limit, the variance in 
the baryonic-stellar mass relation is reasonably small, so it is mostly 
free of stochasticity effects.  We test this more thoroughly below.

To fully probe to GOODS depth, we include results from the D5 simulation
in order to fill out the low end of the mass function; using the 64 star
particle criterion on D5 allows us to consider galaxies down to $\lgmstar
\geq 8.76$.  The normalization of the G6 mass function is lower than that
of the D5 by about 40\% at stellar masses below $\lgmstar = 10.3$, likely
the result of overmerging owing to the lower spatial resolution in G6.

\begin{figure*}[ht]
\epsscale{1.00}
\plotone{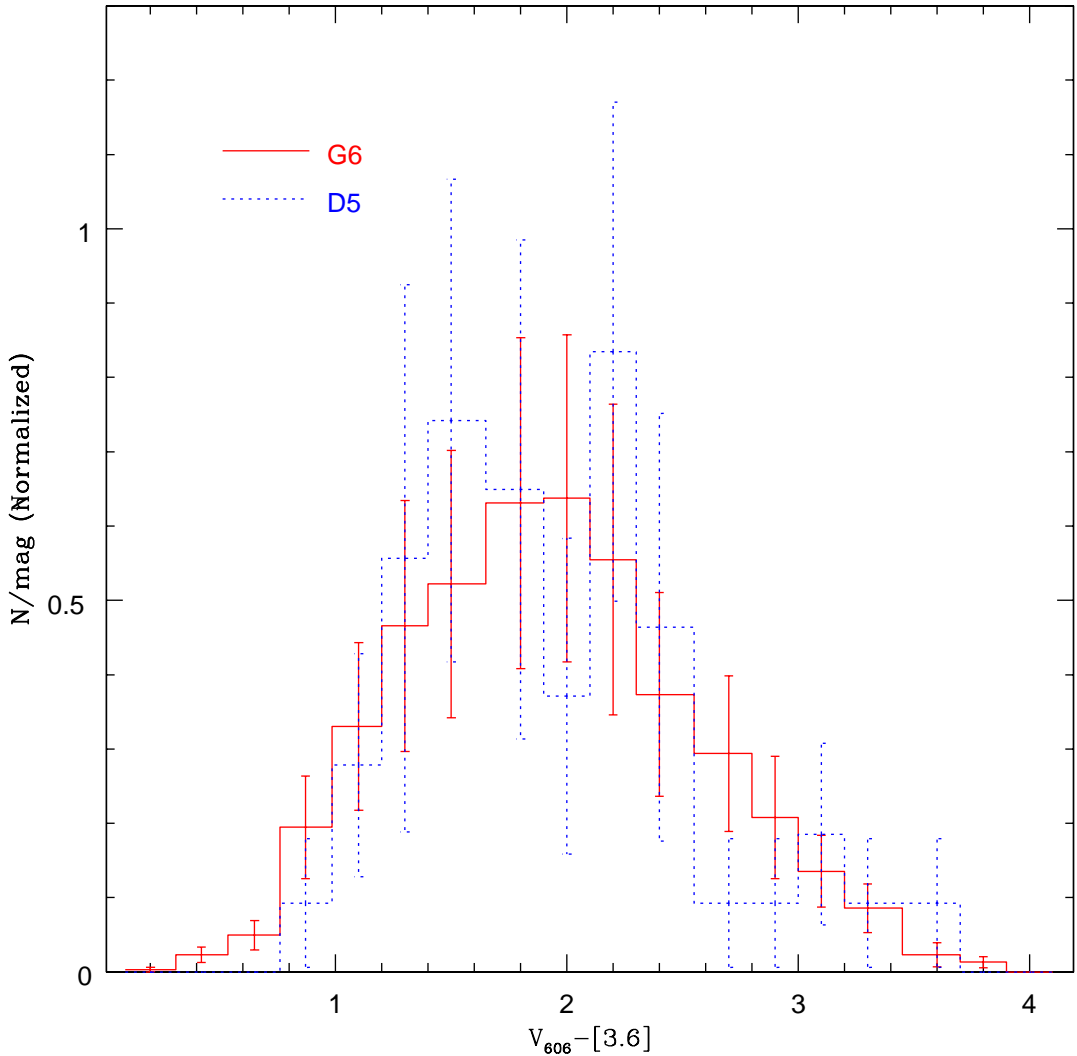}
\caption[]{Histograms of $V_{606} - [3.6]$ colors of simulated B-dropouts with optical 
magnitudes in the range $22 < [3.6] < 24.25$ from G6 (solid, red) and D5 (dotted, blue),
normalized to unit area.  Errors are from jackknife resampling (\S~\ref{sec:lumDensity}).  
Dust reddening has been applied via our fiducial metallicity-derived prescription.
The agreement between galaxies' colors in the high-resolution and low-resolution 
simulations indicates that our stellar mass cut removes galaxies whose star formation
histories are not resolved in the low-resolution simulation.}
\label{colorHist}
\end{figure*}

\parname{Color resolution} A galaxy's color is determined mainly by when
its stars formed.  In a lower resolution simulation, a galaxy remains
unresolved for longer, so that its star formation is delayed.
The galaxy will then be bluer than it should be, and may even experience
an unphysical burst of star formation just as the density threshold for
star formation is exceeded (M. Fardal, private communication).  Indeed,
we find that globally at $z=4$, G6 galaxies have roughly twice the
birthrate (i.e. star formation per unit stellar mass) as D5 galaxies,
indicating that for the general galaxy population in G6, this effect
could be significant.

Does this offset in birthrates propagate into GOODS-observable quantities?
One measure of the birthrate is the $V_{606} - [3.6]$ color, which 
probes the relative amount of light from young, OB-type stars to later-type 
(A and later) stars, thus constraining the ratio of the amount of recent 
star formation to the amount from previous events.  Thus, to answer
our question we constructed histograms of reddened galaxy colors in
observed $V_{606} - [3.6]$ for the
two simulations, shown in Figure~\ref{colorHist}.  The plot shows only
those galaxies whose observed IRAC 3.6$\mu$ fluxes lie in the range $22 <
[3.6] < 24.25$ (corresponding roughly to stellar masses in the range
$9.7 < \lgmstar < 10.5$) and which satisfy the GOODS $z = 4$ LBG color
and magnitude cuts (cf.\ \S~\ref{basicprops}).  There may be a
slight excess at the blue end of the G6 histogram, although this may
result from an underpopulated tail for the D5 distribution owing to its
small volume.  Overall, however, there is remarkable agreement given
the different birthrates; a Kolmogorov-Smirnov test of the two color
distributions yields a 99.2\% probability that they are drawn from the
same distribution.  This suggests that the G6 simulated birthrates and
colors are generally insensitive to resolution effects for our choice
of a 64-star particle cut and GOODS selection.

\parname{Group Finder} The choice of group finder used to identify
galaxies in the simulated data could in principle affect the overall
properties of the selected sample.  Typically, however, the baryonic
clumps representing galaxies are much better separated than dark matter
halos in such a way that unique identification is not difficult.
We compared LBG number counts in the D5 and G6 simulations at $z=3$
to the results of~\citet{N1,N1erratum}, who used an entirely different
group finder on the same simulations.  In the G6 and D5 simulations
with no reddening, they found 12,402 and 202 galaxies that consisted
of at least 32 particles and satisfied the~\citet{Steidel2003} color
cuts, yielding comoving source densities of -1.91 and -2.28 in units
log(N/$h^{-3}$Mpc$^3$).  For the same sample definition and using
our group finder, we extracted 11,626 and 244 LBGs, corresponding to
source densities of -1.93 and -2.19.  These differences are not easy to
interpret, but as they are small we proceed while bearing in mind that
the choice of group finder may introduce a $\approx 10\%$ uncertainty
into volume-averaged predictions; this uncertainty is small compared
to, for instance, uncertainties in our reddening prescription.

\section{PROPERTIES OF THE $z=4$ SAMPLE} \label{basicprops}

\subsection{Sample Definition}

In order to lend concreteness and relevance to our simulated sample,
we study those galaxies that satisfy the GOODS $z \sim 4$ LBG
color selection criteria given in~\citet{Giavalisco2}:
\begin{eqnarray*} 
(B_{435} - V_{606}) \geq 1.2 + 1.4 \times (V_{606}
- z_{850})\; \wedge\; \\
(B_{435} - V_{606}) \geq 1.2\; \wedge\;
(V_{606} - z_{850}) \leq 1.2.  
\end{eqnarray*} 
Unless otherwise stated, we
combine these color cuts with a magnitude cut at $z_{850} < 26.5$,
which is roughly the $10 \sigma$ detection limit for point sources in
$z_{850}$~\citep{Giavalisco1}.  Additionally, unless otherwise stated we
apply dust reddening to our sample using our fiducial metallicity-derived
$E(B-V)$ distribution.  In this section we discuss how these criteria
confine the properties of the sample under investigation.

\begin{figure*}[ht]
\epsscale{1.00}
\plotone{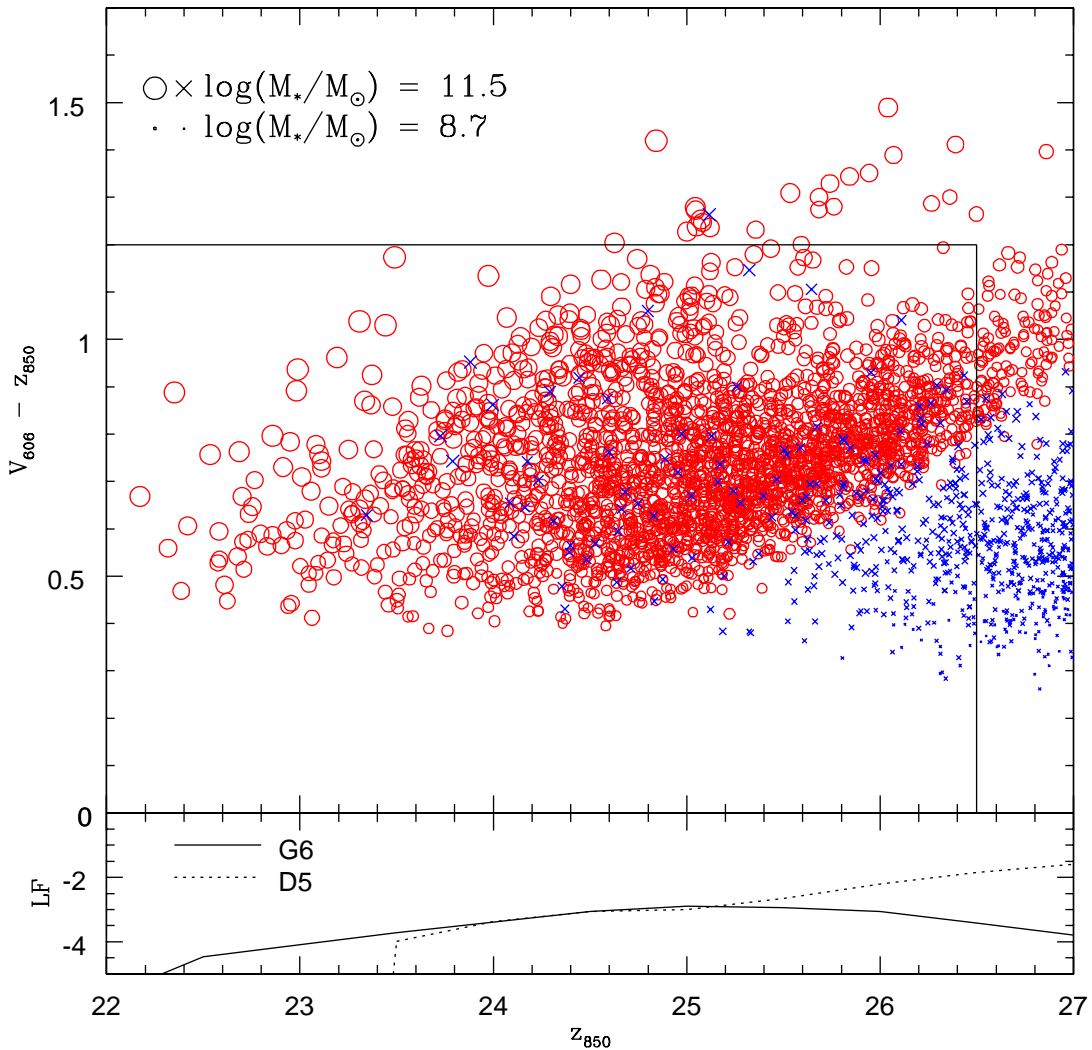}
\caption[]{(top) GOODS color-magnitude diagram of resolved G6 (red circles) 
and D5 (blue crosses) galaxies at z=4.  Point sizes scale linearly with $\lgmstar$ 
as indicated.  The heavy lines indicate the B-dropout color- and brightness cuts.
(bottom) The $z_{850}$ LFs for the G6 (solid) and D5 (dotted) simulations
in units of log(N mag$^{-1}$ $h^3$ Mpc$^{-3}$).  The B-dropout color 
and magnitude cuts miss low-mass galaxies owing to faintness and massive 
galaxies owing to redness.}
\label{goods_cm}
\end{figure*}

The B-dropout color cuts require that LBGs be red in $B_{435} - V_{606}$
and blue in $V_{606} - z_{850}$.  Inspection revealed that all of
the simulated galaxies satisfy the redness cut in $B_{435} - V_{606}$.
This results directly from the~\citet{Madau1} prescription for attenuation
by the IGM, so it is not a direct prediction of our simulation.  From now on,
we ignore the $B_{435} - V_{606}$ color selection criterion.  The upper
panel of Figure~\ref{goods_cm} shows the simulated sample, before
applying color and detection cuts, in the $V_{606} - z_{850}$
vs. $z_{850}$ color-magnitude space.  The point sizes are scaled by
stellar mass while the G6 galaxies are red and the D5 galaxies are blue.
Heavy lines indicate the GOODS cuts.  The magnitude cut excludes most of
the low-mass galaxies and a few of the massive ones while the color cut
independently excludes only a few massive galaxies.  The slope of the discontinuity
between the G6 and D5 samples indicates the line of constant
stellar mass---at a given stellar mass, galaxies that are brighter are
also bluer since both effects result from an increased star formation
rate or a younger stellar population.  The bottom panel plots the LF
for each simulation in units of log(N mag$^{-1}$ $h^3$ Mpc$^{-3}$).
The smooth connection between the two samples where the G6 turns over owing
to our stellar mass minimum gives further evidence that the G6 galaxies
are resolved down to that mass limit.  

\begin{figure*}[ht]
\epsscale{1.0}
\plotone{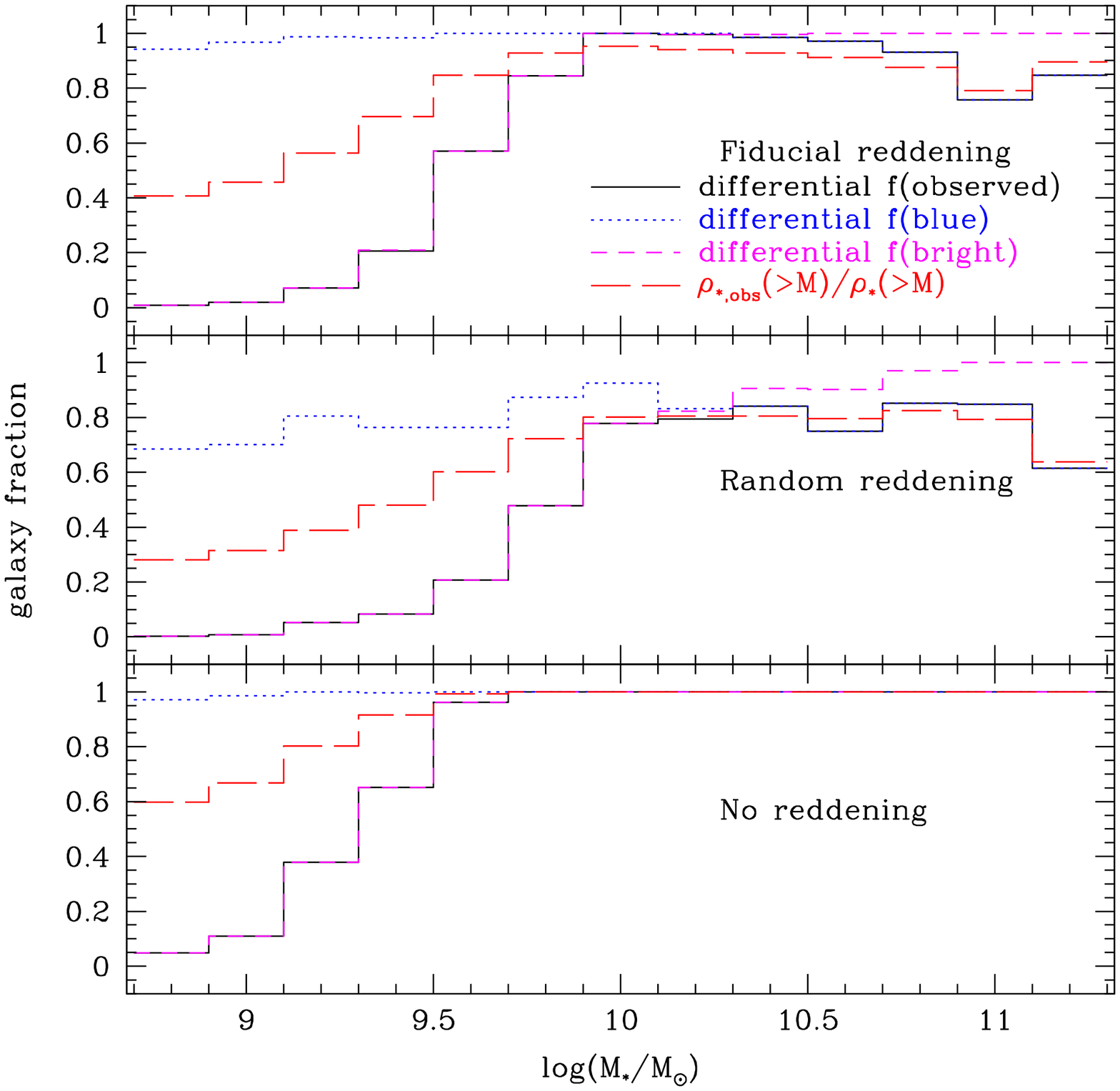}
\caption[]{The mass selection function of LBGs in the simulated GOODS sample for three
different \ebv~distributions: fiducial (top), random (middle) and dust-free (bottom).  
In each plot the blue, dotted line gives the fraction of galaxies in that 
mass bin that are bluer than $V_{606} - z_{850} = 1.2$; the magenta, 
short-dashed line gives the fraction that are brighter than 
$z_{850} = 26.5$; the black, solid line gives the fraction that are
both bright and blue; and the red, long-dashed line gives the fraction of 
stellar mass density in galaxies at or above that stellar mass that would be 
observable in GOODS.  For any reasonable intrinsic~\rd, the B-dropout 
selection criteria remove low mass ($\lgmstar < 9.5$) galaxies owing to 
faintness and massive galaxies owing to dust reddening.}
\label{mass_completeness}
\end{figure*}

Figure~\ref{mass_completeness} shows the mass selection function of the
simulated GOODS sample for the fiducial and random \rd{s} as well as
a dust-free case (top, middle, and bottom panels, respectively); thus,
the top panel is a different way of looking at Figure~\ref{goods_cm}.
Since none of the simulated galaxies are excluded by the cut in $B_{435}
- V_{606}$, we consider only what fraction of the galaxies, as a function
of stellar mass, are bluer than 1.2 in $V_{606} - z_{850}$ (dotted blue
line), brighter than 26.5 in $z_{850}$ (magenta short-dashed line),
or both bright and blue---that is, observable (solid black line).
Not surprisingly, low-mass galaxies are missed owing to faintness while
some massive galaxies are missed owing to redness.  Interestingly, both of these
reasonable $E(B-V)$ distributions predict that the selection functions
reach their maximum around $10^{10} \; \msun$ with a fairly steep dropoff
to lower masses.  For the fiducial case, there is additionally a shallow
decline toward higher masses owing to the higher reddening experienced by
more massive, metal-rich galaxies.  In either case, 10--20\% of galaxies
with stellar mass larger than $10^{10} \; \msun$ are missed because dust
causes them to be too red in $V_{606} - z_{850}$.  This effect accounts
for much of the difference between reddened and unreddened rest-frame
UV \lf{s} at the bright end of Figure~\ref{lumfunc1}.

The red long-dashed line shows, as a function of stellar mass, the
cumulative fraction of stellar mass density in galaxies at or above a given
stellar mass that is selected by the GOODS color and magnitude cuts
and is in principle observable.  This can be thought of as a normalized
convolution of the mass function $\phi(m)$ with the selection function
$f_{\mbox{\scriptsize obs}}(\mbox{M}_*)$:

\begin{equation}
{\rho_{*,\mbox{\scriptsize obs}}(>\mbox{M}_*) \over \rho_*(>\mbox{M}_*)} = {\int^{\mbox{M}_{\mbox{\scriptsize max}}}_{\mbox{M}_*}\phi(m) f_{\mbox{\scriptsize obs}}(m) m \ud m \over
\int^{\mbox{M}_{\mbox{\scriptsize max}}}_{\mbox{M}_*}\phi(m) m \ud m}
\end{equation}

Several points are of interest.  First, the simulations predict that,
even in an ideal universe with no dust, no more than 60\% of the total
stellar mass density residing in galaxies more massive than the resolution 
limit at $\lgmstar = 8.76$  would be observable owing largely to 
low-mass galaxies being too faint.  In fact, this is an upper 
limit on the actual completeness over all galaxies since the 
simulated galaxy mass function is likely prolific at masses below
the resolution limit.  Second, once dust is added the observed fraction
falls to between 30\% and 40\%, depending on the $E(B-V)$ distribution.
Third, some fraction of real galaxies is always unobservable owing to
blending with stars or other galaxies, further suppressing the fraction of
stellar mass density observable; this effect is routinely accounted for via Monte
Carlo simulations.  Finally, it is interesting to repeat the exercise
for the cumulative star formation density by simply replacing $m$ with
$\dot{m}$ in the above integral, and comparing.  The total cumulative
fractions of stellar mass density observed are 41, 28, and 60\% for the fiducial,
random, and dust-free reddening distributions, respectively; the total
cumulative star formation fractions observed are 51, 36, and 69\% for the
same distributions.  Thus, the observed galaxies may contain a somewhat
higher (by $\approx10\%$) fraction of the star formation density than
the stellar mass density.  This is expected since the brightness cut in
$z_{850}$ corresponds to a rest-frame UV selection, which is expected to
trace the star formation rate better than the stellar mass.  However,
the small magnitude of the effect indicates that such a selection does
not exclude the bulk of the stellar mass.

Previously,~\citet{N2004c} compared the total density of stellar mass
in the G6 simulation to observations at $z\sim3$ and found that the
simulation includes roughly twice as much stellar material as is
observed at high redshift if the simulation is normalized so as to 
agree with observations at low redshift.  They speculated that the 
missed stellar mass may be hidden in massive galaxies that are too red 
to be observed.  Figure~\ref{mass_completeness} continues this line 
of inquiry.  Our findings are in qualitative agreement in that 
we show how no more than half of the stellar mass is likely 
to be observable via the Lyman break technique.  We confirm 
that a portion of the missed stellar mass is hidden in 
massive, dusty galaxies, although the number of such galaxies is
sensitive to the amount of dust extinction applied.  However, 
as mentioned we also find that much---and possibly most---of the missed 
stellar mass at this redshift is hidden in numerous low-mass galaxies 
that are simply too faint to be observable rather than in massive red 
or ``red and dead" systems.
 
It is interesting to compare our completeness results to those of
semi-analytic models.  \citet{Idzi1} used a semi-analytic model to
predict that the B-dropout selection criteria select less than 50\% of
the stellar mass density in galaxies more massive than $\log(M_*/\msun) > 9.57$.
For the same range of stellar mass, we find that the B-dropout criteria
select 82\% and 58\% of the stellar mass density for the cases of fiducial
and random reddening, respectively.  The difference almost certainly
arises from the very different ways in which the two models treat star
formation.  For example, our LBGs are generally more massive than those
of~\citet{Idzi1}, and the trend between stellar mass and rest-frame UV
flux is much stronger in our sample than in theirs (cf.\ their Figure 2).
Our results qualitatively agree with theirs in the sense that most of
the missed stellar mass results from the magnitude limit at $z_{850} <
26.5$ rather than the color cuts.

The fact that a reasonable fraction of massive galaxies is missed by the
$B$-dropout color cuts raises the question of whether different selection
criteria could isolate a larger---or at least different---portion
of the high-redshift galaxy population.  \citet{daddi2004} proposed a
two-color criterion for isolating high-redshift galaxies that is intended
to include highly obscured star-forming galaxies as well as massive
galaxies with old stellar populations.  Their criterion is $RJL \equiv
(J - [3.6]) - 1.4(R - J) \geq 0$.  This color cut excludes \textit{none}
of the resolved galaxies in our simulations for redshifts $z < 4.7$
(by $z \approx 4.7$, the IGM substantially suppresses
flux in observed $R$, rendering the galaxies too red in observed $R-J$).
Hence, our simulations suggest that this may be an effective way to select
galaxies that are missed by the Lyman dropout technique as long as
samples are not heavily contaminated by low-redshift interlopers.

Finally, we remark on the expected sample completeness in the IRAC bands.  
While the rest-frame UV data probe to $z_{850} = 26.5$, the IRAC data set
achieves rough 5$\sigma$ detection limits of 26.1, 25.5, 23.5, and 23.4 in 
the [3.6], [4.5], [5.8], and [8.0] channels, respectively.  Our simulations 
indicate that with these limits, the fraction of B-dropouts that will be 
detected is 96\%, 82\%, 12\%, and 11\% for the same bands.  Thus, 
only the [3.6] and [4.5] channels reach deep enough to observe a 
majority of galaxies selected via the B-dropout criteria.

\subsection{Luminosity Density of the Universe} \label{sec:lumDensity}

\begin{figure*}[ht]
\epsscale{1.0}
\plotone{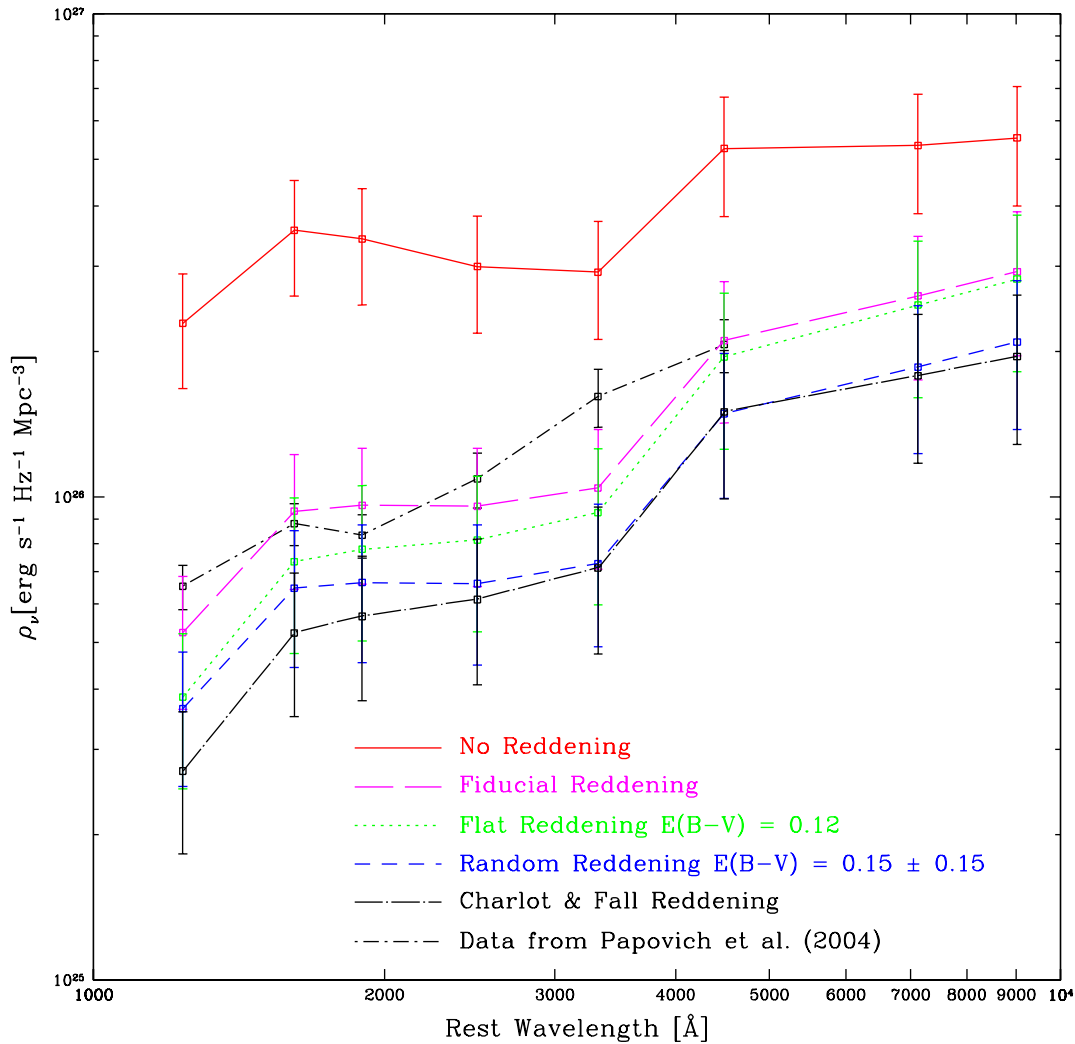}
\caption[]{The luminosity density of the universe for B-dropouts at $z\sim4$ 
in the rest-frame UV to optical bands. The solid, red line gives the $\lnu$ curve 
assuming no dust reddening; the magenta, long-dashed line assumes our fiducial 
reddening prescription; the green, dotted line assumes a uniform reddening for 
all galaxies; the blue, short-dashed line assumes a Gaussian distribution 
of \ebv~values (see text); the black dot-long-dashed curve assumes
the~\citet{CF2000} reddening curve; and the black dot-short-dashed curve
gives the observations from~\citet{Papovich1}.  The fiducial reddening
prescription produces broad agreement with the observations while
showing a small but interesting discrepancy in the observed H-band (see 
text for discussion).}
\label{lumDensity}
\end{figure*}

With the expansive baseline in wavelength afforded by GOODS, we
can construct a coarse average spectrum of LBGs by summing the
flux densities of all the galaxies through each photometric band.
The resulting SED is known as the ``luminosity density" $\lnu$, and can
be used to study the average properties of the galaxy population. For
example,~\citet{Papovich1} used this method to suggest that the universe's
stellar mass density increases by 33\% between $z = 4$ and $z = 3$,
and~\citet{Rudnick2003} used it to infer that the density of stellar mass
in massive galaxies grows by $10\times$ between $z = 3$ and $z = 0$.

We compute $\lnu$ following the method of~\citet{Papovich1}.  We consider
only those galaxies that satisfy the B-dropout color cuts, the brightness
cut at $z_{850} < 26.5$, and an additional brightness cut at $i_{775}
< m^*(i_{775}) + 1$, where $m^*(i_{775}) = 25.74$ is derived from the
value used in~\citeauthor{Papovich1} (\citeyear{Papovich1}, which in
turn was derived from~\citealt{Giavalisco2}) by redshifting from $z=3.9$
to $z=4$.  The flux densities of the selected galaxies in all relevant
photometric bands are converted to luminosities, added, and divided by
the simulation volume to produce the luminosity density $\lnu$.

We compared the luminosity densities of the two simulations computed
in this way as a further resolution check on the G6 simulation,
using only galaxies whose stellar masses were in the range $9.7 <
\log(M_*/\msun) < 10.5$ since this range is resolved by both simulations
(cf.\ Figure~\ref{mass_function}).  We found that, for these galaxies, the
D5 luminosity density is 81\%--94\% that of G6 at this redshift for the
various bands.  This is smaller than the uncertainty in each simulation
owing to Poisson noise and cosmic variance, which we derive below, so
we ignore this offset.  Since neither of our simulations completely
covers the mass range of interest, we compute the total \ld~by summing
the independent contributions from G6 galaxies with $\log(M_*/\msun)
\geq 9.7$ and D5 galaxies with $\log(M_*/\msun) < 9.7$.

We estimate uncertainty in $\lnu$ using the jackknife
method \citep{Lupton1993,Zehavi2002,Weinberg2004}.  In general, the uncertainty
in estimates of volume-averaged cosmological quantities such as the
luminosity density $\lnu$ receives independent contributions from
Poisson noise and cosmic variance.  One simple way to account for both
effects is to divide up the sampled region into smaller subvolumes of
comparable size, estimate $\lnu$ in each volume, and take the variance
in these estimates as the overall uncertainty.  As~\citet{Zehavi2002}
have noted, this method tends to overestimate the error owing to cosmic
variance on scales that are larger than the size of the subsamples.
The jackknife method utilizes subsamples whose sizes are similar to that of
the full sample.  In this method, each subsample is obtained by excluding
a small region (such as one octant of the simulation) from the full 
sample in turn and estimating $\lnu$ in the remaining sample.  The 
full uncertainty is then readily obtained from the formula:
\begin{equation}
\sigma^2(\lnu) = \frac{N-1}{N} \sum_{i = 1}^N (\rho_{\nu,i} - \langle\lnu\rangle)^2
\end{equation}
where $N$ indicates the number of subsamples.  We obtain the uncertainty in 
the $\lnu$ for each simulation by computing jackknife errors over octant 
subsamples in this way and, when combining G6 and D5, we add the 
respective errors in quadrature (this is justified since the two 
simulations are run with different random number seeds and therefore 
provide independent measurements of the same quantity).

Figure~\ref{lumDensity} shows the result.  The solid red line
is obtained in a dust-free universe; the purple long-dashed, green
dotted, and blue short-dashed lines use the~\citet{Calz1} reddening
curve with our fiducial, flat, and random $E(B-V)$ distributions,
respectively; the black dot-long-dashed curve uses the~\citet{CF2000}
reddening prescription; and the black dot-short-dashed curve gives the
data of~\citet{Papovich1}.  The~\citet{Calz1} curve is widely used,
but the~\citet{CF2000} law is not yet as common despite its sound
physical motivation.  In its simplified form, it uses a $\lambda^{-0.7}$
reddening curve and attempts to account for the dusty environments around
young stars by assuming that light from stars that are younger than the
maximum age of their birth clouds (which we take to be $10^7$ years)
has an optical depth to dust scattering that is greater than the
optical depth for the older stars by some factor (which we take to be 3).
For this curve, we took the normalization $A_V$ to be such that the
``old" stars for each galaxy suffer the same extinction in rest-frame
V as the entire population in our fiducial metallicity-derived case;
the young stars are then at three times this optical depth.

It is clear that the fiducial reddening prescription produces broad
agreement with the data from the rest-frame UV into the rest-frame
optical.  It is also not too different from the results of the other
\rd{s}.  In addition, for reasonable amounts of dust extinction, the
\emph{shape} of $\lnu$ is not sensitive to the reddening prescription:
all of the simulated curves show relatively flat UV and optical continua
with a strong break at 4000 \AA.  We emphasize that the agreement between
the data and the $\lnu$ from our fiducial \rd~has not been enforced by
hand; the fiducial distribution has only been calibrated against nearby
late-type galaxies from SDSS.

While all of the simulated data points are affected by dust extinction,
the points at 1200 \AA\ (observed $V_{606}$) are suppressed by an
additional $0.4 \pm 0.1$ magnitudes ($\approx 25\%$) owing to IGM
absorption, as noted in \S~\ref{subsec:reddening}.  Meanwhile,
the simulated points at 7200 and 9000 \AA\ (observed [3.6] and [4.5])
are determined primarily by the stellar masses of the simulated LBGs
with little dependence on dust attenuation. They are reasonably robust,
and straightforwardly testable, predictions of our simulation.

It is interesting that the simulated \ld~curve has a strong break at 4000
\AA~that is not visible in the data.  The 4000 \AA~break results from
a large number of metal absorption features at wavelengths $\lambda <
4000$ \AA~acting in concert with the Balmer break at 3646 \AA.  As it
occurs only in the spectra of relatively cool stars, it is visible only
in stellar populations older than $\sim100$ Myr and is widely used as a
characteristic age indicator.  The stellar populations in our galaxies
have characteristic ages of 300 Myr (Figure~\ref{phys_phys}); therefore
a break is expected.  Although star formation histories of LBGs are
currently poorly constrained by observations, the work of~\citet{Shapley1}
suggests that a substantial population of galaxies whose mean stellar
ages are old enough to show a 4000 \AA~break should be visible by
$z\sim4$; thus, one might expect to see this feature in the data.
The ingredients that are most likely to give rise to the discrepancy
are redshift scatter, dust reddening, the simulated star formation
histories, and uncertainty in the data.  As the 4000 \AA~break is an
important feature for characterizing stellar populations, we consider
these possibilities in turn.

Redshift scatter is present in the data since the redshifts of LBGs
are known only to within the width of the selection function resulting
from the color cuts that are used to define the LBG sample.  We find
that the B-dropout color cuts select most galaxies with $3.5 < z <
4.5$ and very few galaxies outside this range; thus, the optical data
in Figure~\ref{lumDensity} can be viewed roughly as a convolution of
the unknown, correct luminosity density curve with a tophat function
of width 400--900 \AA, depending on the observed band.  We tested
whether this could remove the 4000 \AA~break by introducing artificial
scatter of $\pm 0.5$ into the simulated galaxies' redshifts before
applying the B-dropout selection criteria and found that the simulated
luminosity density dropped in normalization but did not change shape.
The normalization drop results from moving galaxies into redshift ranges
where the selection function is lower, and the 4000 \AA~break is unchanged
because the width of the selection function is not broad enough to allow
significant flux from redwards of the break to spill into the observed
H band (rest-frame 3200 \AA).

As always, dust reddening is a possibility.  With enough dust any galaxy's
observed UV continuum can be reddened to the point that the 4000 \AA\
break disappears.  Unfortunately, if we add enough dust to our galaxies
(using the ~\citealt{Calz1} law with a reasonable intrinsic \rd) that 
the shape of the simulated \ld~curve approaches the observed shape, its 
normalization drops more than an order of magnitude below the observed 
normalization owing to too many galaxies being reddened out of the
observed sample.  Thus for this explanation to be viable, the real LBGs
must be significantly more luminous and dustier than our simulated sample;
observations in the [3.6] and [4.5] bands should constrain this possibility.
An alternative check on the normalization is provided by the rest-frame UV
\lf~(\S~\ref{subsec:lumfuncs}), which shows broad agreement given our
fiducial \rd.  Figure~\ref{lumDensity} thus tells us that the simulated
4000 \AA~break is insensitive to reddening over a variety of \rd{s} and
two well-calibrated reddening curves if we wish to preserve agreement
between the observed and simulated rest-frame UV \lf{s}.  Hence, we do
not think that reddening is the main cause although it may contribute.

Our currently favored explanation for the ``missing" 4000 \AA~break
is the uncertainty in the data of~\citet{Papovich1}.  Their sample was
culled from $\approx 50$ square arcminutes of data (or $\approx 5\%$ of
the comoving volume in our simulation) owing to the limited areal coverage
of the \textit{VLT}/ISAAC data, and their bootstrap uncertainties do
not account for cosmic variance (note, however, that the break was not
observed in their U-dropouts at $z\sim3$ either).  More insight into
this problem could be gained if the \ld~measurement were reproduced
with the full GOODS data set including deep imaging in the critical
\textit{VLT}/ISAAC H-band.

\subsection{Luminosity Functions}
\label{subsec:lumfuncs}

\begin{figure*}[ht]
\epsscale{1.0}
\plotone{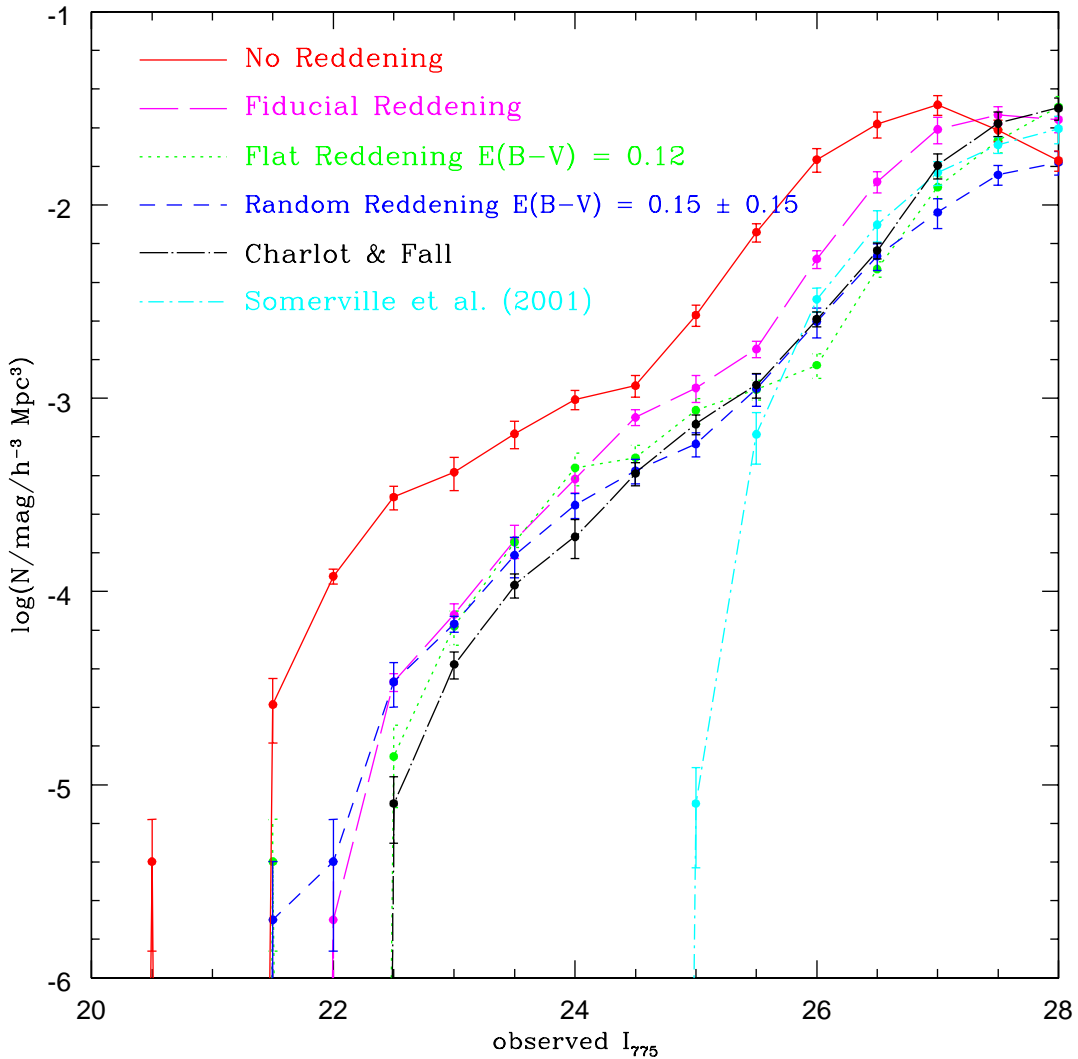}
\caption[]{The rest-frame UV LF of B-dropouts assuming various 
different \rd{s}.  The solid red curve assumes no reddening; the dotted
green curve assumes a uniform reddening; the short-dashed blue curve assumes
a Gaussian distribution of \ebv~values (see text); the long-dashed magenta curve
assumes our fiducial reddening prescription; the cyan dot-short-dashed curve
assumes the~\rd~of~\citet{SPF1}; the black dot-long-dashed curve assumes 
the~\citet{CF2000} reddening curve.  For a variety of reasonable \rd{s}, the
simulated LF varies by $\leq 2\times$ from the curve that assumes our
fiducial \rd.}
\label{lumfunc1}
\end{figure*}

In this section we present rest-frame UV and optical luminosity functions
for the simulated B-dropout sample.  Several studies of the luminosity
functions of LBGs in these simulations have already been published.
\citet{N1,N1erratum} found that the simulated \lf{s}~at $z\sim3$
in observed $V$ and $R$ reproduce the observed \lf~at the bright end
reasonably well while exhibiting steep slopes $\alpha \sim -2$ at the
faint end.  \citet{Night1} performed a similar study for LBGs at $z$ =
4 -- 6 and found good agreement between simulated and observed \lf{s}~in
$i'$ and $z'$ provided that \citet{Calz1} reddening was assumed with
\ebv~= 0.15 -- 0.30.  Additionally, they found a steep slope $\alpha
\sim -2$ at the faint end of the simulated \lf.  At the risk of some
overlap with their work, we will present a complete discussion of
the LBGs in these simulations under a single set of assumptions such
as our mass cuts and assumed \rd.  In order to create a single \lf,
we combine the results of the D5 and G6 simulations following the same
method as described in \S~\ref{sec:lumDensity}.

\parname{Reddening} Figure~\ref{lumfunc1} shows the simulated $i_{775}$
luminosity function for a number different \rd{s}.  The galaxies
used in creating this plot satisfy the GOODS color cuts but do not
necessarily satisfy the GOODS detection limit $z_{850} < 26.5$.  The
solid red line shows the \lf~derived if dust is absent so that \ebv=0.
The magenta long-dashed, green dotted, and blue short-dashed lines
show the \lf~under the assumptions of a fiducial, flat, and random~\rd,
respectively (\S~\ref{simdescript}); and the cyan dot-short-dashed
line assumes that the reddening varies with star formation rate in the way
described by~\citet{SPF1}.  All of these curves use the~\citet{Calz1}
reddening curve once the reddening \ebv~for each galaxy is chosen.
The black long-dash-dotted line uses the~\citet{CF2000} reddening curve
as described in \S~\ref{sec:lumDensity}.

We briefly consider the various reddened curves in turn.  Inspection
reveals slight differences between the results from the flat, random,
and fiducial $E(B-V)$ distributions.  Our fiducial $E(B-V)$ distribution
naturally produces a weak positive correlation between stellar mass and
reddening that is absent in the random distribution, preferentially
reddening massive galaxies out of the sample (cf.\ Figure~\ref{mass_completeness}).  
For the same reason, at the faint end the random \rd~causes more
galaxies to be reddened out of the sample than the fiducial distribution.
The random distribution allows a few more bright galaxies to have low
dust content than either the flat or fiducial \rd{s}.  The curve from
the flat distribution is similar to the others except at the bright end,
where, once again, the redder intrinsic colors cause these galaxies
to be preferentially reddened out of the sample if high reddening is
enforced.  Despite these minor differences, it is clear that
the random, fiducial, and flat \rd{s} produce qualitatively similar
results.  The curve derived from the~\citet{SPF1} \rd~reddens rapidly
star-forming galaxies out of our sample far too effectively;
it is clearly not appropriate for our simulated galaxies.  The curve derived from
the~\citet{CF2000} reddening prescription has almost exactly the same
shape as our fiducial curve but is fainter by roughly 0.5 magnitude;
this is the result of adding extra extinction to the youngest stars
in each galaxy.  It would not be difficult to modify the parameters
we used for the~\citet{CF2000} law to obtain better agreement, but as
little insight would be gained from this exercise, we do
not do so.

As with the observed mass fraction, the dominant source of uncertainty
in these predictions is clearly the unknown form of the intrinsic \rd~and
reddening curve in high-redshift galaxies.  However, the important point
is that the overall shape of the simulated \lf~is fairly insensitive to
the details of the reddening procedure such that reasonable distributions
agree to within a factor of 2.  This gives us further confidence that
our fiducial \rd~will yield useful predictions.

\begin{figure*}[ht]
\epsscale{1.0}
\plotone{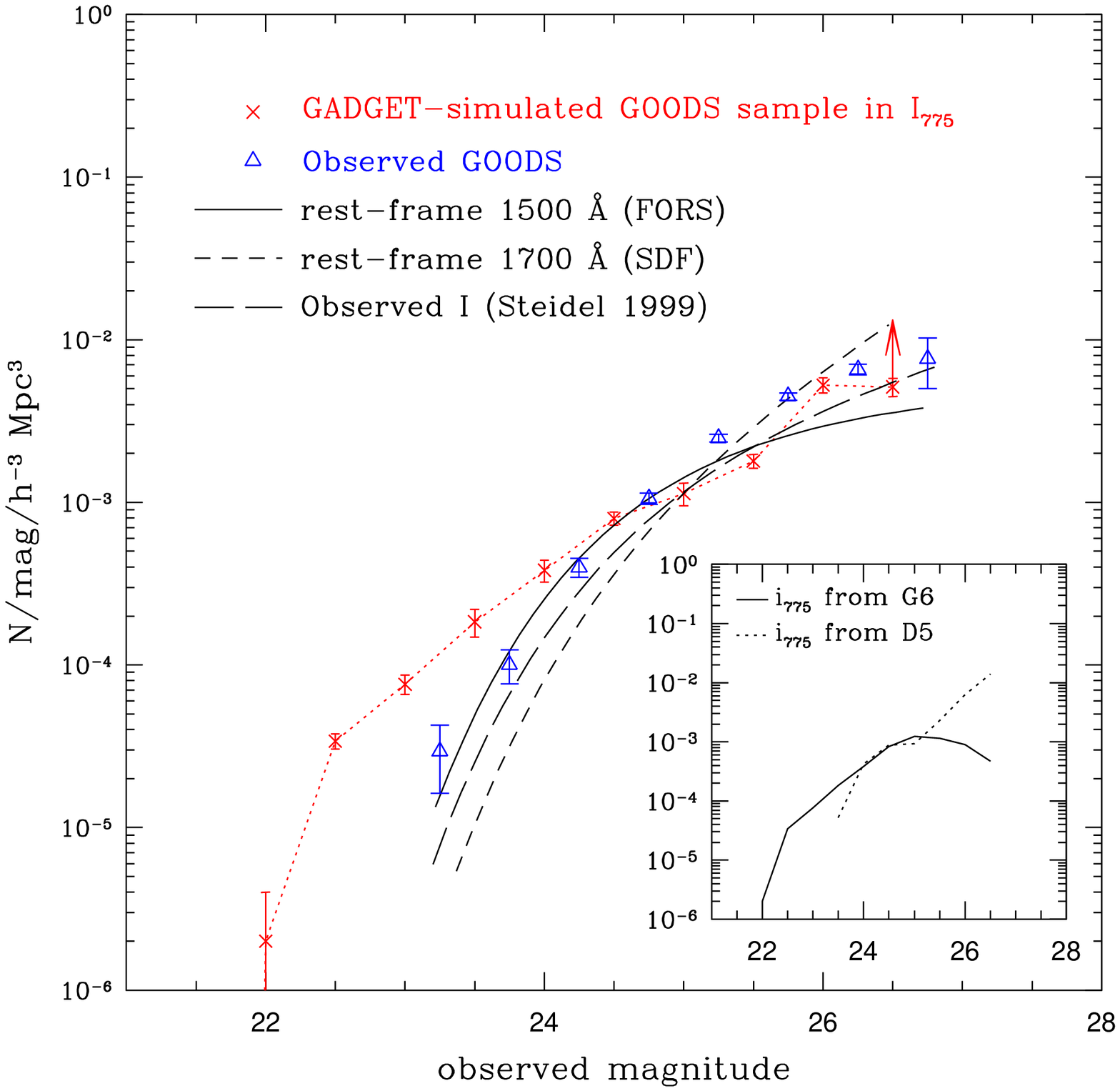}
\caption[]{(large) The simulated rest-frame UV LF of B-dropouts (assuming
our fiducial reddening prescription) compared with several published 
curves and the GOODS data. Red crosses indicate the simulated LF in
observed $i_{775}$; the arrow points to the value of the LF at that 
point if the brightness cut in $z_{850}$ is relaxed.  
Errors in the simulated plot are from jackknife 
resampling.  LFs from the FORS survey~\citep{Gabasch1} at rest-frame 
1500 \AA, the Subaru Deep Field~\citep{Ouchi1} at rest-frame 1700 \AA, 
and Steidel et al.~\citep{Steidel1} in observed $I$ are indicated by 
solid, short-dashed, and long-dashed curves, respectively, and were 
derived from the literature as described in the text.  The GOODS B-dropout 
sample of~\citet{Giavalisco2} in observed $i_{775}$ is indicated by blue 
triangles.
(inset) The LFs of the D5 and G6 simulations, 
which were combined to produce the full curve as described in the text.
The simulated LF assuming our fiducial \rd~produces broad agreement with 
available constraints, with a possible excess at the bright end (see 
text for discussion).} 
\label{lf_uv}
\end{figure*}

\parname{Comparison to Observations} Figure~\ref{lf_uv} shows a
comparison of the simulated $i_{775}$ \lf~using the fiducial \rd~with
various observational results.  The red crosses with jackknife errors
indicate the simulated GOODS sample, while the arrow at the faint end 
points to the value obtained by relaxing the brightness cut $z_{850} < 26.5$.
The plot also includes observed LFs from the FORS survey~\citep{Gabasch1} 
at rest-frame 1500 \AA, the Subaru Deep Field~\citep{Ouchi1} at 
rest-frame 1700 \AA, and Steidel et al.~\citep{Steidel1} in observed 
$I$; these are indicated by solid, short-dashed, and long-dashed curves, 
respectively.  The FORS LF was constructed at $z = 4$ using the fitting 
function given by Equation 1 and the values given by Table 4 in~\citet{Gabasch1}.
The Steidel et al.\ $z\sim4$ LF was constructed from the updated Schechter 
parameters for U-dropouts ($z\sim3$) given in~\citet{AS1} and following 
the method described in~\citet{Steidel1}: we used a distance modulus to 
shift $m_*$ from $z=3.04$ to $z=4.0$, multiplied the normalization 
$\phi_*$ by 80\%, and left $\alpha$ constant, yielding 
($\phi_*$, $m_*$, $\alpha$) = ($3.52\times10^{-3}$, 25.02, -1.57).
The Subaru Deep Field LF is the LF for $BRi$-LBGs from Table 4
of~\citet{Ouchi1}, where we have used their derived faint-end 
slope $\alpha=-2.2$.  Note that because the effective wavelengths of the
bands used in these surveys as well as the mean redshifts of their LBG 
samples differ, offsets of 0.1--0.3 magnitudes between their reported 
LFs do not necessarily indicate disagreements.

The observed and theoretical GOODS LFs both flatten out
for magnitudes fainter than 26.  By contrast, the magnitude-unlimited
GOODS \lf~seems to turn over at a fainter magnitude.  In fact, this
turnover owes to the conservative cut in stellar mass: The more we
relax this restriction, the fainter the magnitude at which the rest-frame UV
\lf~turns over.  Hence, our simulation is predicting an extremely steep
intrinsic faint-end slope ($\alpha \sim -2$) that cannot be observed
in B-dropout samples of GOODS depth.

LBGs brighter than 25th in $i_{775}$ also seem to be overproduced in
the simulation.  This is analogous to the bright-end excess noted in the
same simulation at $z=2$ by~\citet{N2005a}.  The excess could result 
from any of four possibilities: (1) incorrect prescription for 
dust reddening; (2) lack of truncation of star formation, owing 
to e.g. feedback from black hole growth; (3) overmerging; and (4) 
incorrect feedback model for star formation.  We consider each of 
these possibilities in turn, as they provide interesting
insights into high-redshift galaxies.

The first possibility is consistent with the idea, noted
by~\citet{Shapley1} and others, that dust extinction correlates with
luminosity, and hence in our model, star formation rate.  Clearly, if
our assumed extinction scaled more strongly with star formation rate,
closer to that in~\citet{SPF1}, then the discrepancy would be alleviated.
Evidence in support of this possibility is provided by~\citet{daddi2004}:
For their actively star-forming $BzK$ sample at $z>\sim1.4$ with
$K<20$, they find a median star formation rate of $200 \; \smyr$
and a median reddening of $E(B-V) \sim 0.4$.  Indeed,~\citet{N2005b} 
found that the simulations can reproduce the comoving number density of 
Extremely Red Objects (EROs) at $z =$ 1--2 if a uniform reddening of 
$E(B-V) = 0.4$ is assumed for the entire simulated sample.  Using our 
fiducial \rd, simulated galaxies at $z=4$ with star formation rates 
between 150 and 250 $\smyr$ suffer a significantly weaker median reddening 
of roughly 0.15.  Additionally, we note that some of the most massive, rapidly 
star-forming galaxies could correspond to extremely dusty starbursts that are 
observed as sub-millimeter sources~\citep{Chapman1}, as we will consider 
further in \S~\ref{sec:high_sfrs}.  Recent observations suggest that the
effects of dust on these and other massive galaxies ($\lgmstar > 11$) may 
not even follow simple foreground screen attenuation laws such as that 
of~\citet{Calz1} (\citealt{Chapman2}; Papovich et al.\ 2005, in preparation).  
In this case, the~\citet{CF2000} reddening curve is almost certainly more 
appropriate; unfortunately, observations to date have not constrained the 
parameters needed to use it at high redshift.

The second possibility is that our simulation may fail to account for
some mechanism that suppresses or truncates star formation in real
galaxies.  In this case, the bright-end excess is an early sign of the
difficulties that are commonly seen in cosmological simulations and
SAMs~\citep{Somerville2004,N2005a}.  Upcoming work incorporating AGN feedback
may help with this~\citep{DiMatteo2005,springel2005a,springel2005b}.

The third possibility is overmerging owing to poor spatial resolution.
Figure~\ref{mass_function} shows that galaxies with stellar masses
$\lgmstar < 10.25$ are slightly underrepresented in the G6 simulation
relative to the D5.
Owing to its poorer spatial resolution, galaxies that form in the
G6 are more extended than in D5, rendering them more susceptible to
tidal disruption and subsequent accretion by more massive galaxies.
Massive galaxies in G6 then grow more rapidly than they would in
a higher-resolution simulation, leading in principle to a brighter
characteristic magnitude M$^*$.  The impact of overmerging on the
bright end of a rest-frame UV LF can be characterized by inspecting the
galaxy mass function at the massive end.  Figure~\ref{mass_function}
shows that, up to the maximum galaxy mass produced in the smaller,
higher-resolution D5 simulation, the mass functions of the D5 and G6
simulations agree quite well.  Together with the evidence that the birthrates in
the G6 are numerically resolved (Figure~\ref{colorHist}), this gives us
confidence that overmerging is not the dominant source of the discrepancy.

The final possibility is an incorrect model for supernova feedback.
In particular, our simulation's prescription for kinetic feedback
endows all galactic outflows with a common velocity, 484 km s$^{-1}$.
This may be an oversimplification.  For example, \citet{Martin2005} found
that outflow velocities in a sample of 18 ultraluminous infrared galaxies
correlate well with star formation rate, and suggested that this may result
in winds more effectively suppressing star formation in massive galaxies.

In summary, our study of the rest-frame UV LF confirms the previously cited 
studies in the following respects: (1) the simulated and observed LFs show broad 
agreement when a reasonable amount of dust extinction is assumed; (2) the intrinsic 
faint-end slope of the simulated LF is $\sim -2$; and (3) an excess at the bright end 
of the simulated LF suggests that more work is required to understand the effects 
of star formation and AGN feedback as well as dust reddening on massive galaxies.
The fact that our work reproduces these results despite our having employed a 
different group finder and a more physically-motivated prescription for dust 
reddening suggests that these results are robust to the analysis procedure.

\begin{figure*}[ht]
\epsscale{1.00}
\plotone{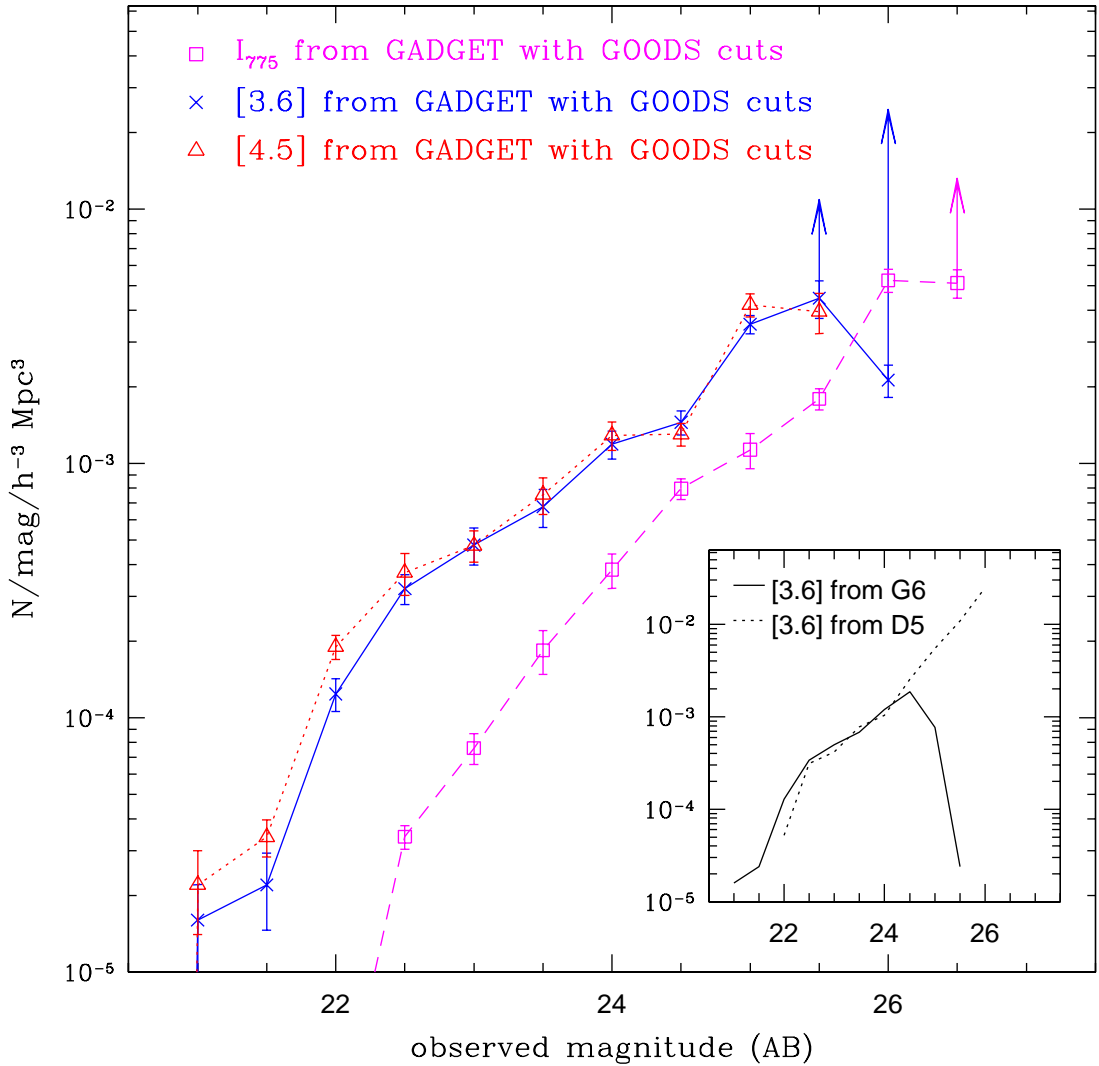}
\caption[]{(large) The simulated rest-frame UV and optical LFs of B-dropouts.
The arrows point to the values of the LF if 
the brightness cut in $z_{850}$ is relaxed.  Errors are from jackknife resampling.
(inset) The \lf{s} of the D5 and G6 simulations in observed [3.6], which were 
combined to produce the top curve as described in the text.  The rest-frame 
optical LF is very similar to the rest-frame optical LF, shifted roughly 1--2 
magnitudes brighter.}
\label{lf_irac}
\end{figure*}

Figure~\ref{lf_irac} shows the rest-frame optical \lf{s} and compares
them with the rest-frame UV \lf.  The observed [3.6] and [4.5] \lf{s} are
very similar, suggesting that measurements of stellar mass may readily
be made in the more sensitive [3.6] channel.  Since LBGs tend to be
1--2 magnitudes brighter in rest-frame optical than in rest-frame UV,
the rest-frame optical \lf~is essentially the same as the rest-frame
UV \lf, shifted 1--2 magnitudes brighter.  This is consistent with
the trend, discussed below, that LBGs are simply the most massive of
the early galaxies, and that their \sfr{s} correlate with their masses.
A closer look suggests that the rest-frame UV \lf~appears to be slightly
steeper than its optical counterpart, consistent with the tendency
for more massive galaxies to be redder.  The simulations predict that
the faint end of the intrinsic \lf~is essentially a power-law down
to much fainter masses than GOODS can observe; the turnover in the
intrinsic \lf~at [3.6] $\approx 26.5$ owes to the 64 star particle
mass resolution cut.  Similarly, the turnover in the observed \lf~at
[3.6] $\approx 25$ results directly from the GOODS magnitude cut at
$z_{850} < 26.5$ and translates into a mass cut at $\lgmstar \approx 9.5$
(Figure~\ref{phys_phot}; see also Figure~\ref{mass_completeness}).

\section{COMPARING PHYSICAL AND PHOTOMETRIC PROPERTIES OF LBGs} \label{sec:detailedprops}

In this section we focus on correlations between the properties of
the simulated sample, comparing to the data wherever possible.  Unless
otherwise noted, we employ our fiducial, metallicity-derived $E(B-V)$
distribution with the~\citet{Calz1} reddening curve.  We occasionally 
compare the physical properties of our simulated sample at $z=4$ with the 
properties derived from observations at lower redshift $z\sim3$ since LBGs 
at this epoch are well-studied.  This may not be a valid comparison since 
some evolution may occur in the population of star-forming galaxies between 
these two epochs~\citep[e.g.,][]{Papovich1}.  On the whole, however, the two 
populations are expected to be very similar~\citep{Steidel1}; thus, we proceed 
while noting that factor of 2 differences do not necessarily suggest a failing 
of the simulation.

\subsection{Physical Properties}
\label{physprops}
\begin{figure*}[ht]
\epsscale{1.00}
\plotone{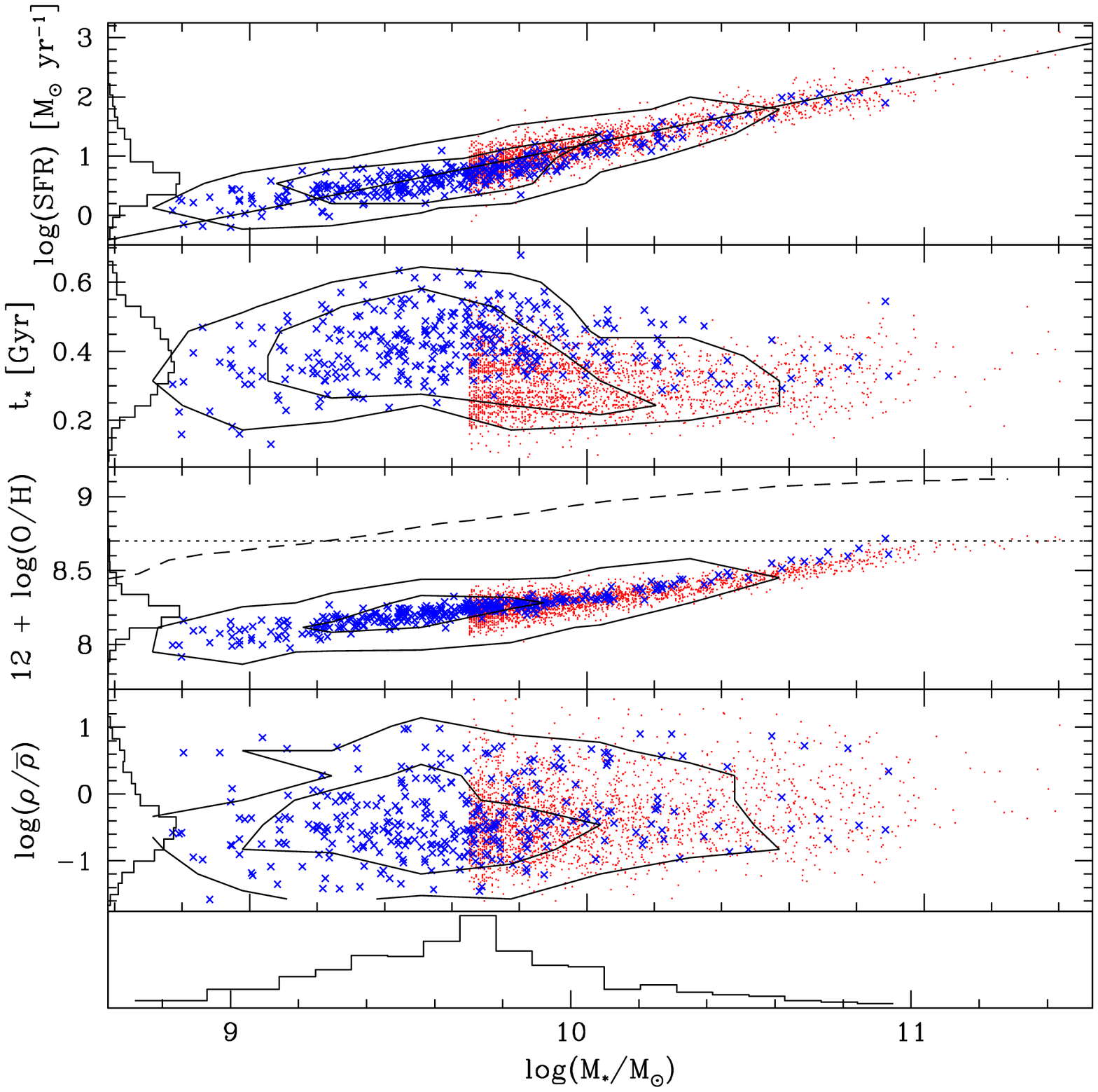}
\caption[]{Physical properties of B-dropouts versus stellar mass.  (top) instantaneous
star formation rate; (second) median age of the star particles; (third) mean metallicity
of the star particles; (fourth) local density of observable stellar mass, relative to the mean.
Red points and blue crosses denote G6 and D5 galaxies, respectively.  The contours 
enclose 63\% and 95\% of the simulated galaxies.  The histograms and contours are 
constructed by volume-weighting the contributions of the D5 and G6 samples as described 
in the text.  Note the strong correlation between star formation rate and stellar mass.}  
\label{phys_phys}
\end{figure*}

Figure~\ref{phys_phys} shows a number of physical correlations in
addition to the basic properties of the simulated GOODS $z\sim4$ sample.
Red points and blue crosses denote G6 and D5 galaxies, respectively.
The break in the G6 sample at $\lgmstar = 9.64$ owes to our 64 star
particle cut.  The histograms are constructed by considering only G6
galaxies more massive and D5 galaxies less massive than $\lgmstar =
9.7$ and weighting the contribution from the D5 galaxies by the ratio
of the simulation volumes.  The contours were constructed using the same
method and define regions enclosing 63\% and 95\% of the galaxies.
For each galaxy, 
$\lgmstar$ is the sum of the masses of the star particles; 
the star formation rate is the instantaneous star formation rate;  
$t_*$ is the median age of its star particles; 
$12 + \log(O/H)$ is the mean metallicity of its star particles;
and $\rho/\overline{\rho}$
is the local space density of stellar mass in B-dropouts (roughly
equivalent to an optical luminosity-weighted number density), normalized
by the mean over the simulation volume.  We compared instantaneous
star formation rates with star formation rates obtained by averaging
over the last 100 Myr and found that they are within a factor of 2
(\S~\ref{sec:high_sfrs}), in agreement with \citet{Weinberg2002}.

\parname{Stellar Mass} The stellar masses of our sample fall predominantly
in the range $\lgmstar = $ 9--10 with a peak in the range 9.5--9.8 and
a significant tail out to $\lgmstar \approx 11$.  The falloff at lower
masses owes entirely to the brightness cut at $z_{850} < 26.5$ (cf.\
Figures~\ref{goods_cm} and~\ref{mass_function}).  Overall, this range
is slightly more massive than the sample from the semi-analytic model
of~\citet[Figure 2]{Idzi1}; in particular, those semi-analytic models
do not reproduce our high-mass tail.  \citet{Papovich2} used population
synthesis techniques to infer the properties of the stellar populations
of 33 LBGs in the HDF-N between $z\sim 2$ and $z\sim 3.5$ and found a
typical stellar mass of $10^{10} \msun$.  \citet{Shapley1} used similar
techniques to infer the properties of 74 fairly bright LBGs at $z\sim3$
and found masses in the range $\lgmstar = $ 9--11 with a median at $1.2
\times 10^{10} h^{-2} \msun$.  While these measurements are somewhat
more massive than the median in Figure~\ref{phys_phys}, we note that
both of these samples are biased towards brighter and thus more massive
galaxies than our simulated GOODS sample.  Also, given that~\citet{Papovich1} 
report a $\sim33\%$ buildup of stellar mass between $z=4$ and $z=3$, it
is reasonable to suppose that galaxies at $z=3$ might be more somewhat more
massive than galaxies at $z=4$.  Thus, we conclude that our LBG model seems 
to be able to match the distribution of observed stellar masses.  

\parname{Star Formation Rates} The bulk of the observed galaxies are
forming stars at rates of 1--10 $\msun \mbox{ yr}^{-1}$.  A small active
population is forming stars at aggressive rates of $100\sim 1000 \; \smyr$
with the median instantaneous star formation rate of the ten most rapidly
star-forming galaxies being $\sim 450 \; \smyr$.

Our models produce a tight correlation between stellar mass and star formation 
rate, as noted elsewhere in hydrodynamic 
simulations~\citep[e.g.,][]{Dave99,Weinberg2002,N1}.  A fit to this
relation for our simulations gives
\begin{equation} \label{eqn:sfrmass}
\log(\dot{M}_*/\msun) \approx  1.14\lgmstar - 10.2,
\end{equation}
which is shown in the plot.
As mentioned in \S~\ref{resolution}, a comparison of the mass
functions (Figure~\ref{mass_function}) and the birthrates (not shown)
of the D5 and G6 galaxies suggests that the G6 galaxies may be suffering
slightly from overmerging, but we believe that the star formation rates in the G6
simulation cannot be overpredicted by more than a factor of 2, which is
comparable to the intrinsic scatter.  Figure~\ref{phys_phys} thus predicts
that a few galaxies with instantaneous star formation rates of 500--1000
$\msun \mbox{ yr}^{-1}$ may fall within the GOODS field.  According to our
prescription for dust reddening, they should all be observable as LBGs; however,
considering observational evidence that the most massive galaxies may not
obey the~\citet{Calz1} reddening curve (\citealt{Chapman2}; Papovich 2005, in
preparation), this may not be the case in reality.  We will 
return to these massive star formers in \S~\ref{sec:high_sfrs}.  For 
now, the strong correlation between stellar mass and star formation rate
implies that LBGs observed at $z=4$ are predominantly massive galaxies.

The scatter in the plot of instantaneous star formation rate versus
stellar mass can be regarded as a measure of the star formation rate
``burstiness" as a function of stellar mass.  For example, galaxies with
stellar mass of $10^{10} \; \msun$ have star formation rates between 10
and 30 $\msun \; \mbox{yr}^{-1}$.  This means that such a galaxy will,
on average, tend to form 10 $\msun \; \mbox{yr}^{-1}$ although it will
sometimes experience bursts of up to 30 $\msun \; \mbox{yr}^{-1}$; if
the simulation produced more dramatic bursts than this then there would
be more spread in the plot.  Note, however, that the simulation does
not adequately resolve merger-driven starbursts (e.g. Hernquist \& Mihos
1995; Mihos \& Hernquist 1996; Springel et al. 2005b).  For those galaxies
undergoing mergers, the instantaneous star formation rate reported by
the simulation may be an underestimate even though the total stellar
mass formed by the end of the merger event is reliable.
Relatedly, \citet{Weinberg2002} notes that there could be merger events
between galaxies not massive enough to be resolved in the simulation,
which cause them briefly to be luminous enough to satisfy the B-dropout
magnitude cuts.  However, the broad agreement between the simulated
and observed rest-frame UV \lf{s} suggests that the bulk of the LBG
population can be accounted for without resorting to such phenomena.

The lower end of the distribution of star formation rates is reasonably
consistent with semi-analytic results~\citep{SPF1,Idzi1}.  However,
the space density of more rapidly star-forming galaxies 10--100 $\smyr$
is higher in our results, no doubt owing to the drastically different
star formation and reddening prescriptions involved.  Observationally
inferred LBG star formation rates are in reasonable agreement with
our results~\citep{Shapley1,Papovich2}.  The histogram of \sfr{s}
in~\citet{Shapley1} seems to show a higher space density of galaxies
with \sfr{s} greater than $10\; \smyr$, but since this sample was
hand-selected to span the range of possible properties of LBG stellar
populations rather than to satisfy an unbiased selection criterion, such
an excess of galaxies exhibiting more extreme properties is unsurprising.

\parname{Metallicity} The simulated galaxies' stellar metallicities 
correlate tightly with stellar mass.  The simulation computes metallicities 
by allowing the metallicity of a star-forming gas particle to increase 
during each timestep in proportion to its current star formation rate
and a constant yield that is set to the solar metallicity 0.02; as mentioned,
a star particle inherits the metallicity of its parent SPH particle.
Given that a galactic mass-metallicity relation is not assumed 
in this prescription, it is of interest to examine the trend that 
nonetheless results.  Such a trend is required by a growing body of observational
evidence (e.g.,~\citealt{Lequeux1979},~\citealt{GS1987},~\citealt{Tremonti2}); for
a review of this subject see the discussions in~\citet{Tremonti2}.

The dashed line gives the median relation between stellar mass
and gas-phase metallicity observed by~\citet{Tremonti2} in their sample
of $\sim53,000$ nearby ($z\approx 0.1$) star-forming galaxies.
Despite the simple prescription for enrichment, the slope of
the trend in the simulated sample shows excellent agreement with the 
observed low-redshift trend.  Over a range of two orders of magnitude in 
stellar mass, the simulated galaxies show a stellar metallicity that is 
roughly 0.6 dex below the low-redshift value for their respective stellar 
masses and the most massive objects ($\lgmstar \geq 11$) already exhibit 
solar metallicity.  Closer examination suggests that the simulation may not 
produce the flattening observed by~\citet{Tremonti2} at stellar masses greater 
than $\lgmstar = 10.5$.  While such a disagreement may be a failing if 
observed in {\it gas-phase} metallicities at low redshift, it is not 
necessarily a failing for stellar metallicities at high redshift.

Physical processes that could contribute to the observed mass-metallicity
relation include selective removal of metals by supernova-driven winds
(``depletion," first modeled by~\citealt{Larson1974}~and
favored by~\citealt{Tremonti2}), ISM enrichment owing to supernova
feedback~\citep[``astration,"][]{Lequeux1979}, dilution owing to infall of unenriched gas,
and late-time return of relatively unenriched gas to the ISM from evolved, low-mass
stars (``recycling").  Of these processes, depletion contributes in the simulation
via the superwind prescription, astration is built into the enrichment 
prescription, dilution can contribute, and recycling is not accounted 
for.  If astration dominated the trend then the stellar metallicities 
in all simulated galaxies would be expected to be roughly solar
owing to the constant yield assumed by the star-formation prescription.  
If dilution dominated the trend then massive galaxies, which are 
accreting gas much more rapidly than less massive galaxies, would 
be expected to exhibit lower metallicities---the opposite of what 
is seen.  Thus, the simulation's superwind prescription is the 
most likely cause of the stellar mass-metallicity relation, and the 
excellent agreement between the slopes of the observed and simulated 
relations give us further confidence in the superwind prescription 
as well as in the predicted LBG metallicities.

\parname{Median Age} Figure~\ref{phys_phys} (second panel) shows that the
median age of a galaxy's stars is at best weakly correlated with stellar
mass.  This means that the phenomenon of ``downsizing"~\citep{Cowie96},
i.e. that larger galaxies are observed to have an older stellar population
out to $z\sim 1$, is at best weakly evident in our models at $z=4$.
This is a fundamental and testable prediction of our simulations.

The age distribution extends from 200--600~Myr with a median around
350--400~Myr.  This is roughly a factor of 2 older than the population
in the semi-analytic simulations of~\citet{Idzi1}, consistent with the
picture in which LBGs are massive galaxies rather than starbursts.
These values are also roughly a factor of two older than the constraints 
derived from observed LBGs at $z\sim3$ by~\citet{Shapley1}.  However, the 
uncertainties in star formation histories derived via this technique are 
substantial~\citep{Shapley1,Shapley2005}; thus, agreement within a factor 
of two gives us confidence that our models have reasonably realistic star 
formation histories.

\parname{Environment} The fourth panel gives the galaxy density
$\rho/\bar{\rho}$ in each galaxy's local environment.   We define the
density for each galaxy as the local density (relative to the mean)
of stellar mass in the 16 nearest galaxies that would be observable
by GOODS, smoothed with an SPH kernel; this is broadly equivalent to a
number density weighted by rest-frame optical luminosity.  There is a weak
correlation between stellar mass and local matter density in the sense
that the most massive galaxies are in denser environments.  Galaxies in
denser areas also show redder colors and higher star formation rates
(Figure~\ref{phys_phot}), likely arising as a secondary correlation owing
to the tight relationship between stellar mass and star formation rate.
Although there is considerable scatter, such a correlation might be
observable in the GOODS sample.  We note that weighting by stellar mass
(or, equivalently, rest-frame optical flux) is critical; when we compute
densities using unweighted number counts these weak trends disappear
altogether.

\subsection{Physical Versus Photometric Properties}
\begin{figure*}[ht]
\epsscale{1.0}
\plotone{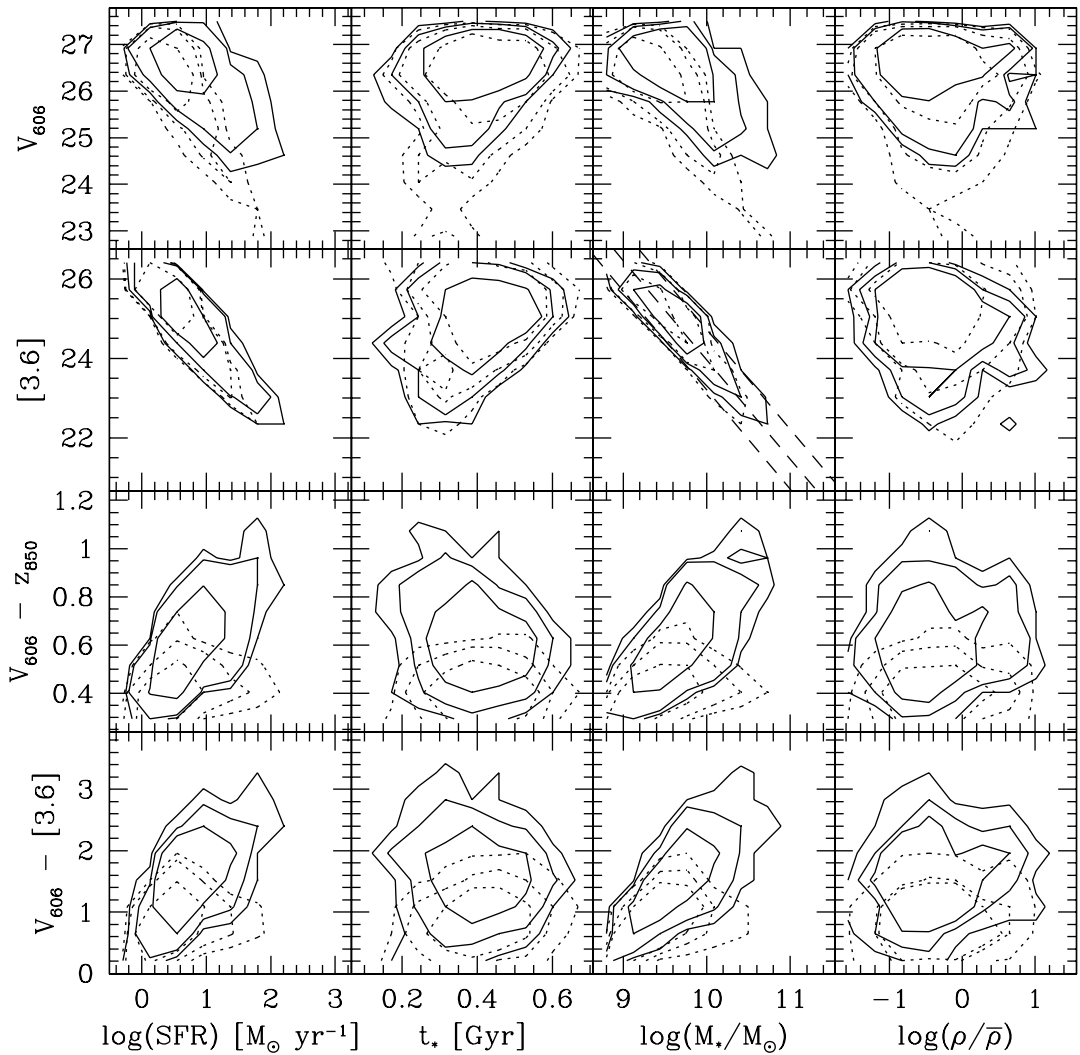}
\caption[]{Physical versus photometric properties of B-dropouts.  Solid 
contours enclose 63\%, 86\%, and 95\% of the galaxies assuming our
fiducial \rd; dotted contours assume a dust-free universe.
Contours are constructed by volume-weighting the contributions of 
the D5 and G6 simulations as described in the text.  The dashed lines
in the plot of stellar mass versus [3.6] flux show constant
mass-to-light ratio corresponding to 1/8 (brightest), 1/4 (middle), and
1/2 (dimmest) of the solar value. Note that star formation rate correlates
more strongly with optical flux than with UV flux, even when reddening 
has not been applied.}
\label{phys_phot}
\end{figure*}

Figure~\ref{phys_phot} compares some physical and photometric
properties of the simulated sample as it would be observed with reddening
(solid contours) and without (dashed contours).
Physical properties include stellar mass, star formation rate, age,
and local galaxy density.  Photometric properties include rest-frame UV
($V_{\rm 606}$) and optical ($[3.6]$) magnitudes, a short-baseline rest-UV
color ($V_{\rm 606}-z_{\rm 850}$) and a long-baseline UV-optical color
($V_{\rm 606}-[3.6]$).  While the GOODS measurements cover all four
IRAC channels, we plot only the [3.6] band here because the [3.6] band
will detect the largest fraction of the B-dropouts (\S~\ref{basicprops}).
We note that the trends observed in [3.6] persist in the redder bands.
Figure~\ref{phys_phot} shows the following interesting properties:

\noindent -- Rest-frame optical data are a better predictor of star
formation rate (SFR) than rest-frame UV data.  This seems surprising at
first, but is expected given the tight stellar mass-star formation rate
correlation in Figure~\ref{phys_phys} along with the fact that reddening
introduces much more scatter into the UV than into optical light.
One may wonder whether rest-frame UV would be a better predictor of
SFR if it were possible to remove the effects of reddening accurately;
the dotted contours in Figure~\ref{phys_phot}
suggest that this would at best allow UV flux to be
an equivalently good predictor, but not better.

\noindent -- The tight correlation between star formation and stellar mass 
also gives rise to a correlation between rest-frame UV flux and stellar 
mass, in qualitative agreement with the trend noted in U-dropouts 
at $z\sim$2--3.5 by~\citet{Papovich2}.  \citet{Shapley2005} do not find 
such a correlation in their UV-selected sample at $z\sim2$.  However,
Papovich et al.\ (2005, in preparation) find a significant correlation 
between stellar mass and star formation rate in their sample of 
near-infrared--selected galaxies at $z \sim$ 1--3.5.  Thus, a careful 
assessment of selection biases is probably necessary for determining 
whether the~\citet{Shapley2005} result is in conflict with the simulations.

\noindent -- More massive galaxies are redder in both the unreddened and
reddened cases, with the trend accentuated in the reddened case
because higher mass galaxies tend to have more dust.
This trend is qualitatively consistent with observations~\citep[e.g.][]{Papovich2}.
It also leads to the counter-intuitive albeit weak trend
that more rapidly star-forming
galaxies are redder.  Unfortunately, this correlation is not
visible in the photometry (substituting a rest-frame UV magnitude for the
star formation rate) if the scattering effects of reddening have not been
removed.

\noindent -- There are no obvious correlations between any photometric measurement
and the galaxies' stellar ages, consistent with the findings
of~\citet{Papovich2} and~\citet{Shapley1} that star formation history
is difficult to constrain from photometry alone.

\noindent -- There is a very strong correlation between stellar mass and rest-frame
optical magnitude, as expected.  This correlation persists in the
longer-wavelength IRAC bands (not shown) but the scatter does not
decrease dramatically.  We note that the correlation between optical flux and 
stellar mass in the G6 simulation at $z=2$ has previously been shown to agree 
well with observations~\citep{N2005b}.

\noindent -- The dashed lines in the figure of $M_*$ vs. [3.6] show lines of constant
mass-to-light ratio corresponding to 1/8 (brightest), 1/4 (middle), and
1/2 (dimmest) of the solar value in this band.  Evidently, the stellar
populations of B-dropouts have roughly 1/4 the solar mass-to-light
ratio in rest-frame optical light.  For comparison, an unreddened 
SSP of the same metallicity with ages between 200
and 600 Myr has rest-frame V-band mass-to-light ratios of 0.12 to 0.24.
The difference between the two ranges owes entirely to reddening: As
the dashed contours in Figure~\ref{phys_phot} reveal, the mass-to-light
ratios for the unreddened galaxies fall neatly within the range given
by the SSPs.  The agreement further indicates that the galaxies' spectra
are not dominated by bursts.

\noindent -- There is a weak trend between color and environment in the sense that
redder galaxies are in more overdense areas, and the trend between stellar
mass and environment does translate into a trend between luminosity and
environment.  It will be interesting to see if stronger correlations are
seen in the data; the ones predicted here will be quite difficult to
pull out of the GOODS sample given additional scatter from photometric
errors (which we have not included).

\subsection{Photometric Properties}
\begin{figure*}[ht]
\epsscale{1.00}
\plotone{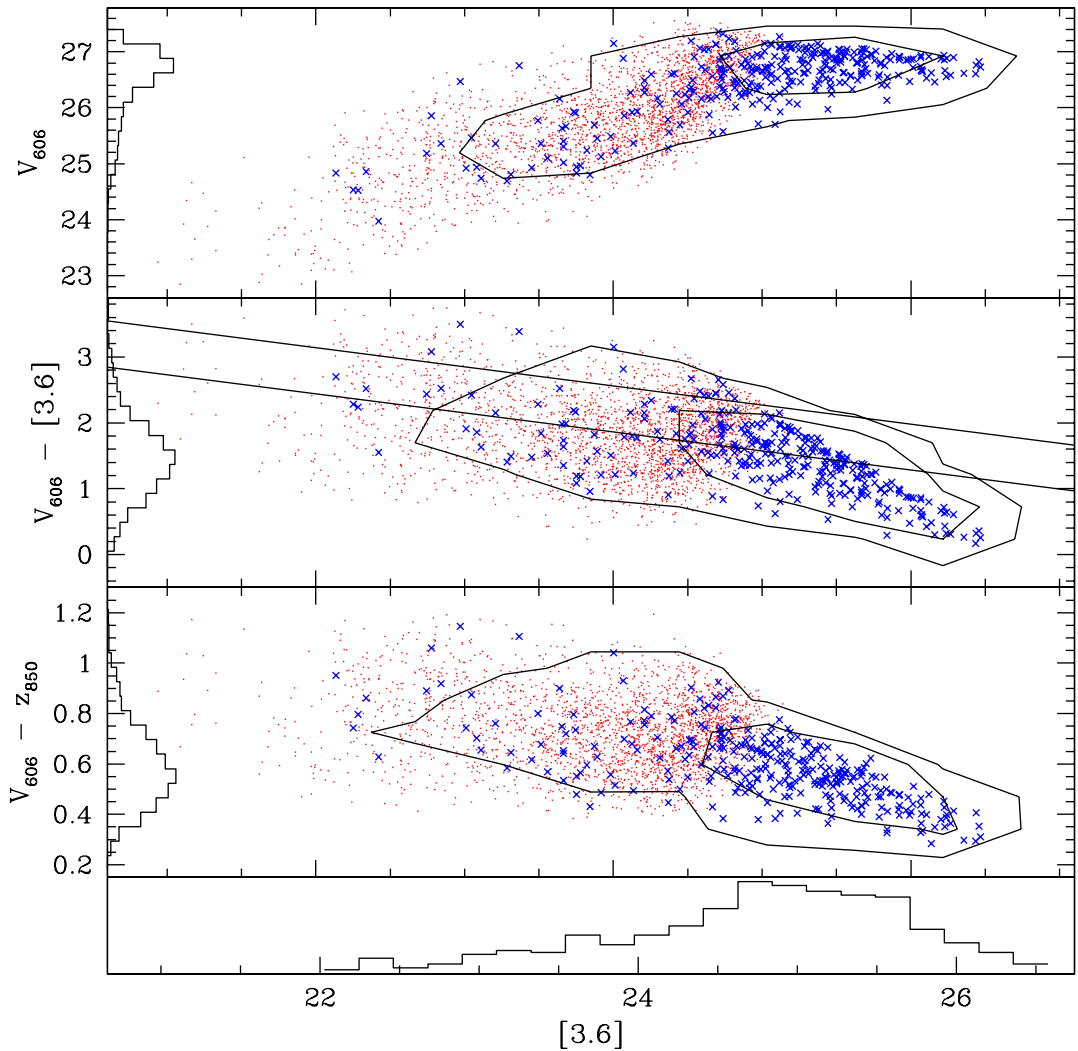}
\caption[]{Photometric properties of B-dropouts versus observed flux in
[3.6].  (top) rest-frame UV; (second) rest-frame UV-optical color; (third) 
rest-frame UV color.  The solid lines give possible color cuts for defining
a high-redshift red sequence (see text).  The contours enclose 63\% and 95\% of the simulated 
galaxies.  The histograms and contours are constructed by volume-weighting 
the contributions of the D5 and G6 samples as described in the text.  Rest-frame
optical and UV flux show a strong correlation.}  
\label{phot_phot}
\end{figure*}

In this section we make predictions for observable photometric
relationships for GOODS galaxies.  Validating these predictions would
lend support to our model for high-redshift galaxies,
while disagreements would hopefully give insights into physical processes
that are absent from the simulations.  Figure~\ref{phot_phot} shows
a short-baseline color, a long-baseline color, and a rest-frame UV
magnitude plotted against a rest-frame optical magnitude ([3.6]).
The following properties are evident:

\noindent -- There is a strong positive correlation between rest-frame UV
and optical flux, consistent with the relationship that the more massive
galaxies are more rapidly star-forming (Figure~\ref{phys_phys}).  This is
a fundamental prediction of our simulations, although the predicted
scatter may be underestimated if starbursts that cannot be resolved by
these simulations are prevalent.

\noindent -- There is a strong correlation between the optical flux and
color for both long-baseline (UV-optical) and short-baseline (UV-UV)
colors.  Since redder colors indicate a lower specific star formation
rate, the simulations are reproducing the important trend in galaxy
evolution noted in~\citet{Papovich2}.

\noindent -- We identify no passively-evolving, ``red and dead" massive
galaxies in our $z=4$ sample.  At lower redshifts, such galaxies
appear in a tight ``red sequence" in color-magnitude diagrams~\citep{Bell1}.
There is as yet no evidence for a red sequence at high redshifts $z
\gtrapprox 1$ although several massive, passively evolving galaxies
have been identified (Papovich et al.\ 2005, in preparation;~\citealt{yan2004};
~\citealt{daddi2005};~\citealt{labbe2005}).  In any case,
the color-magnitude diagrams in Figure~\ref{phot_phot} are clearly not
bimodal.  Indeed, this dearth of passively evolving massive
systems is seen at lower redshifts
in models as well, and highlights an important failure of current galaxy formation
models~\citep{Somerville2004,N2005a}.  Therefore it is of interest to predict
where red sequence galaxies might lie on these plots if they existed at
high redshift.

Using $\sim25,000$ galaxies from the COMBO-17 Survey,~\citet{Bell1}
identified a color criterion for isolating red sequence galaxies and
showed how the population's colors evolve from $z=0.2\rightarrow 1.1$.
One limit on the color of the $z=4$ red sequence is obtained by assuming
that the red sequence does not evolve between $z=4$ and $z=1.1$.
Using the color limit of ~\citet{Bell1} at $z=1.1$, and converting
the rest-frame $U-V$ color and absolute $V$ magnitude to the observed
$V_{606} - [3.6]$ color and observed [3.6] magnitude, gives the upper
line in the plot of [3.6] versus $V_{606} - [3.6]$.  On the other hand,
since roughly 4 Gyr pass between $z=4$ and $z=1.1$ in our cosmology, we
expect that the color of the hypothetical red sequence will be different
at $z=4$.  By assuming that the color evolves linearly with time we obtain
an upper limit to the amount of color evolution experienced by the red
sequence, shown by the lower line.  Both lines succeed in isolating a
sample of fairly massive red galaxies in the reddened sample.  However,
these galaxies are all forming stars at more than 1 $\smyr$ and are
red owing to dust, not low birthrates; if dust is removed then very few
of the simulated galaxies are redder than either cut.
The existence of dust-free massive red galaxies at these epochs, if found,
would be a strong challenge for current models of galaxy formation.

\begin{figure*}[ht]
\epsscale{1.0}
\plotone{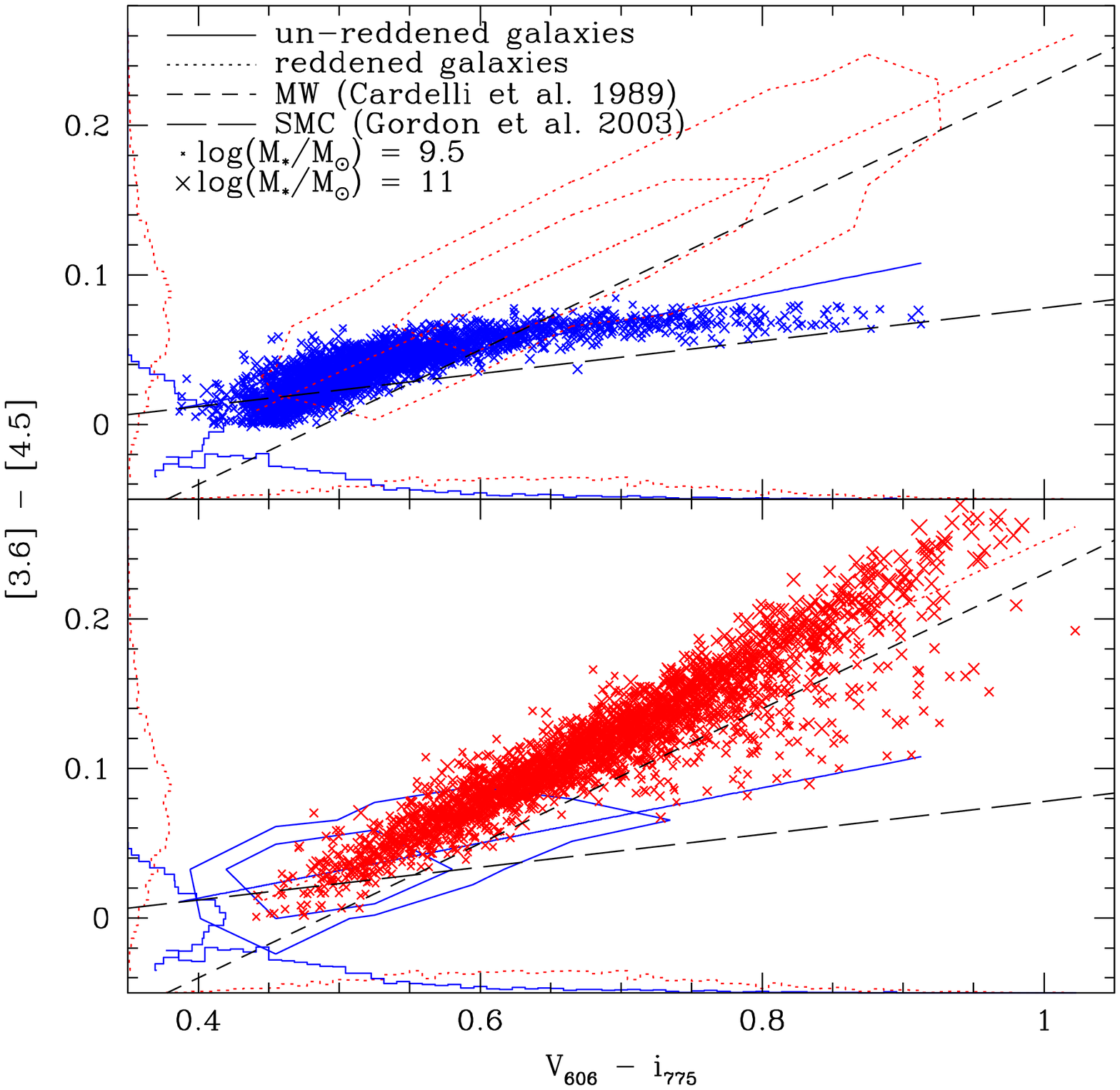}
\caption[]{Rest-frame optical versus UV color of B-dropouts from the G6
simulation assuming fiducial reddening (red) and no 
reddening (blue).  Contours enclose 63\% and 95\% of the galaxies.
Point size scales linearly with $\lgmstar$ as indicated.  
(top) reddened LBGs are given by the contour, unreddened LBGs are given by 
crosses; (bottom) reddened LBGs are given by crosses, unreddened LBGs are
given by the contours.  The solid, dotted, short-dashed, and long-dashed
lines give regression fits to locus of B-dropouts assuming no reddening,
fiducial reddening, a Milk Way reddening curve, and the SMC bar reddening
curve.  The slope of the color-color plot varies with the reddening
curve and the median color varies with the typical amount of dust
in the galaxies.}
\label{cc_dust}
\end{figure*}

Figure~\ref{cc_dust}~illustrates how ACS and IRAC data from GOODS may
be used to constrain both the amount and form of reddening.  We have
produced photometry for two simulated GOODS samples from the G6 simulation, 
one in which there is no dust reddening and one in which there is substantial
dust reddening.  In the top panel of Figure~\ref{cc_dust}, the blue crosses
represent the unreddened LBGs while the red contours enclose 63\% and
95\% of the reddened LBGs; in the bottom panel the red crosses represent
the reddened LBGs while the blue contours enclose 63\% and 95\% of the
unreddened LBGs.  The point sizes are scaled to the logarithm
of the stellar mass as indicated.  The histograms on the
side show the normalized distributions of colors for each sample.
The unreddened sample consists of a dense ``blue clump" along with a
diffuse ``red tail" of low-mass galaxies whose optical color is nearly
independent of UV color.  In the reddened sample the ``blue clump"
has been spread out along the direction of the reddening vector---which
results directly from the reddening curve employed~\citep{Calz1}---and
many of the ``red tail" galaxies have been reddened out of the sample.

Roughly, the median colors can be calibrated to yield the typical
dust extinction.  Furthermore, the slope of the regression fit tells
about the reddening law itself, since different laws will scatter
points along different reddening vectors.  As an illustration, we have
included the regression fit to the reddened sample for the case of the
Milky Way reddening law~\citep{CCM1989}, and the average law for the
stars in the Small Magellanic Cloud (SMC) bar from~\citet{Gordon1}.
In all cases we have used the fiducial \rd~to obtain the rest-frame V
extinctions $A_V$.  While the Calzetti and Milky Way laws yield similar
results, the SMC law produces much more reddening in the rest-frame UV
as compared to the rest-frame optical, yielding a much flatter slope in
this color-color space.  Although the differences in color for the 
different extinction laws are somewhat subtle, such differences may be 
distinguishable in GOODS data and could help constrain the general form 
of reddening at high redshift.  Unfortunately, this technique probably 
would not be effective on an individual galaxy basis owing to photometric 
errors.

\begin{figure*}[ht]
\epsscale{1.0}
\plotone{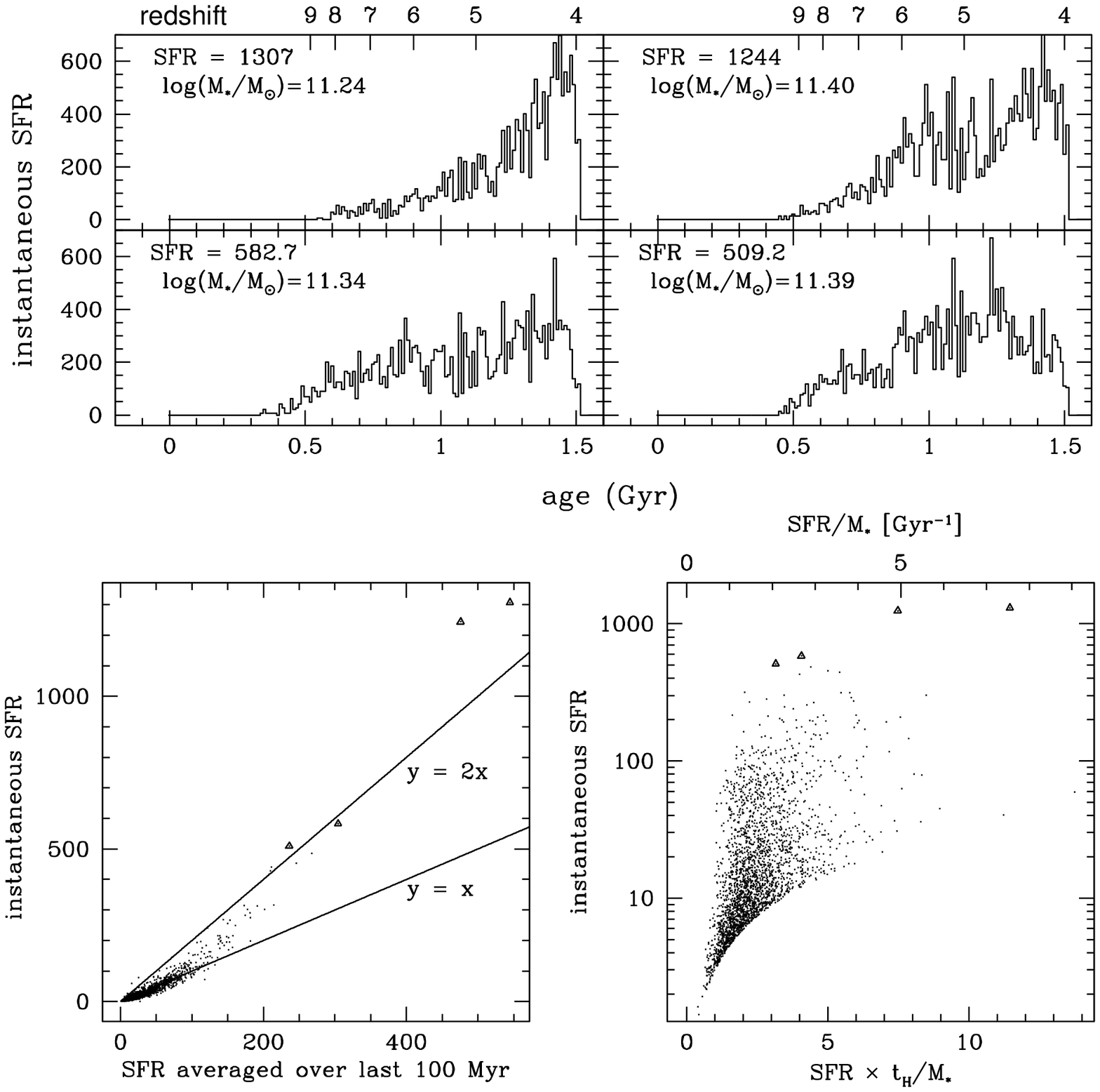}
\caption[]{(top) Star formation histories of the four most rapidly 
star forming galaxies in the simulated samples.  The instantaneous star 
formation rates and stellar masses at $z=4$ are given in the top left; 
the star formation histories are binned in intervals of 10 Myr. 
(bottom left) Instantaneous versus time-averaged star formation rate for 
the GOODS-selected galaxies in the G6 simulation, with the four 
most rapid star formers marked by triangles.  The straight lines give the
location of galaxies whose instantaneous star formation rates are (bottom
line) equal to and (top line) twice the time-averaged rates.  (bottom 
right) Instantaneous \sfr~versus birthrate for the same sources.  The
lower envelope arises from our $10^{9.64} M_{\odot}$ mass resolution 
cut.  These massive, rapid star formers are experiencing at most a 
mild burst of star formation and may be identifiable with high-redshift 
submillimeter galaxies.}
\label{fig:high_sfrs}
\end{figure*}

\subsection{High Star Formation Rate Objects} \label{sec:high_sfrs}

Figure~\ref{phys_phys} shows that the simulated GOODS sample includes
a number of galaxies with instantaneous star formation rates in excess
of 100~$\smyr$, including two in excess of $\sim 1000 \; \smyr$.  As noted
in \S~\ref{subsec:lumfuncs}, these galaxies are part of the bright-end
excess in the rest-UV luminosity function (cf.\ Figure~\ref{lf_uv}), and
may simply represent some failing in the numerics or feedback modeling of
high-redshift galaxies.  However, the intriguing alternative is that these
galaxies are indeed genuine predictions of our model, and are not observed
in the rest-UV because they are in fact much more heavily extinguished than
our assumed dust relation (equation~\ref{eqn:ebv}) predicts.  Indeed,
such a population of heavily extinguished, rapidly star forming galaxies
has been observed at high redshift as sub-millimeter galaxies (SMGs)
using SCUBA.  It is therefore of interest to investigate the physical
properties of these systems in more detail.

The star formation histories of the galaxies with the four
highest instantaneous star formation rates are shown in the top of
Figure~\ref{fig:high_sfrs}, with the instantaneous \sfr~given in
the top left corner and the stellar masses given above each plot.
Despite their prodigious star formation rates, they are not undergoing
significant starbursts; their star formation rates are all increased by
a modest $2-3\times$ over the rate given by equation~\ref{eqn:sfrmass}.
Closer inspection reveals that all of these galaxies reside in highly
overdense regions resembling the cores of collapsing clusters, and owing
to the large number of smaller galaxies in the vicinity, they exhibit a
complex morphology that could be construed as indicating interaction.
However, none of them appear to be the merger of two massive gas-rich
galaxies.  While our simulations cannot resolve the detailed gas dynamics
within each galaxy, it seems unlikely that their star formation is
being driven by dynamical instabilities that funnel gas to a small
central core as in merger simulations (e.g. Barnes \& Hernquist 1991, 1996).
Instead,
it appears likely that star formation is occurring throughout the galaxy,
and the high SFRs are purely a result of the high gas accretion rates
expected for these massive systems.

For more quantitative insight, we compared their instantaneous SFRs
with their 100~Myr-averaged SFRs (bottom left) and their birthrates
(bottom right).  The two most prodigious star formers are forming
more than 1000~$\smyr$, or about three times their 100~Myr averages.
They show birthrates in excess of 5 (i.e., they are forming stars
over 5 times more rapidly than would be needed to acquire their
stellar mass at $z=4$ within a Hubble time).  The other two galaxies
are forming stars at 500--600 $\smyr$, or roughly twice their 100 Myr
averages, and show instantaneous birthrates of $\approx$ 3--4.  In fact,
the instantaneous versus averaged \sfr~plot suggests that the galaxies
whose SFRs exceed $100 \; \smyr$ depart smoothly from the trend set by the
less rapidly star-forming ($\dot{M}_*/\msun < 100$) galaxies.  Hence
these high-SFR galaxies are not anomalies, but a natural extension of
the trends from lower masses and SFRs.

The regime of $\geq 1000\; \smyr$ objects has generally been thought to
be populated by SCUBA sources, which are in turn ultraluminous IR sources
thought to be scaled-up analogs of today's ULIRGs.  Can our high-SFR
galaxies be SCUBA sources?  Unfortunately, very little information is
available on sub-mm galaxies at $z\sim 4$.  However, we can compare
to samples at $z\sim 2-3$, and the results are intriguing.  First,
the space density of our two SFR$>1000 \; \smyr$ galaxies is comparable
to that of $z\sim 2.5$ sub-mm galaxies with $L_{\rm FIR}\ga 10^{13}
h^{-2}L_\odot$~\citep{Chapman2}.  Hence they have broadly similar
star formation rates and number densities,  even if the numbers of
SMGs decreases by a factor of a few from $z\sim2.5\rightarrow 4$
(as seen in Figure 4 of~\citealt{Chapman2}).  Even more curious is
that~\citet{Chapman1} find, for a sample of 12 sub-mm galaxies with
$z\approx 2.2$, the majority show extended radio emission indicative
of a spatially-extended starburst.  This seems more in accord with the
nature of our simulated objects than local ULIRGs that look more like
canonical major mergers with bright central starbursts, although we note
that the merger-driven model of Hopkins et al. (2005a,b) for the
simultaneous origin of starbursts and quasars also predicts a density
of sub-mm galaxies similar to that observed (Hopkins et al. 2005c).

It would clearly be interesting to compare the statistical properties of the 
simulated massive galaxies to observations in more detail.  Although this 
is beyond the scope of the present work, we note that~\citet{N2005b}, who 
used the same SPH simulations, found good agreement between the 
simulated and observed number densities of massive galaxies at $z = $1--3 and
used this finding to argue that hydrodynamic simulations do not suffer from
the ``mass-scale problem."  Since our massive galaxies at $z=4$ must be the 
progenitors of their sample, their work lends further indirect support to the 
suggestion that the our massive galaxies may correspond to an observable 
population such as the SMGs.

In summary, our simulations produce several rapidly star-forming objects
whose properties are broadly in accord with those of SMGs seen with SCUBA.
They are forming stars at a rate that is at most mildly in excess of what
is expected for their stellar mass, indicating that gas infall and not
merger-induced dynamical instabilities is driving their star formation.
For these objects to be SMGs, their dust extinction must lie significantly
above the relation we assume for the bulk of our LBG population
(equation~\ref{eqn:ebv}), but given the ad hoc nature of that relation,
this does not seem far-fetched.  More detailed modeling is required, such
as that done in~\citet{Fardal1}, to see if this hypothesis is viable.

\section{CONCLUSION} \label{sec:conclusion}

We have analyzed the physical and photometric properties of galaxies at
$z=4$ in a cosmological hydrodynamic simulation of a \lcdm~universe.
We employ a novel prescription for modeling dust reddening based on
each simulated galaxy's metallicity, and determine a mass criterion that
robustly resolves the star formation history of these galaxies.  We
examine the properties of galaxies that satisfy the selection criteria
for the $z\sim 4$ B-dropout sample of the GOODS survey.

We subdivide our conclusions into two broad areas: Observed properties
of the GOODS sample, and physical properties of high-redshift galaxies.
For the former, we find:
\begin{itemize}

\item The simulated luminosity density from rest-frame UV through
rest-frame optical bands is in broad agreement with observations.
While the normalization is uncertain to $\sim2\times$ owing to dust
reddening, the {\it shape} of the luminosity density curve is insensitive
to reddening, for reasonable dust models.  We find that the simulated
luminosity density curve includes a conspicuous 4000 \AA~break that is
not seen in observations; whether or not this discrepancy 
is genuine can be tested
with increased areal coverage in the $H$-band, as well as luminosity density
measurements in IRAC bands.

\item The simulated rest-frame UV \lf~is also in broad agreement with the
data.  The turnover at the faint end owes entirely to the magnitude limit;
the intrinsic galaxy population in the simulation has a steep faint-end
slope of $\alpha \sim -2$.  The simulated rest-frame optical \lf~is 
similar to the rest-UV \lf, shifted brighter by 1--2 magnitudes.

\item The simulated rest-UV LF shows an excess of galaxies with $i_{775}<25$
that could indicate a failing of the simulation, but may also represent a
population of highly-obscured, rapidly star-forming galaxies that are
not modeled correctly by our reddening prescription.

\item The GOODS B-dropout selection criteria miss low-mass galaxies owing
to the brightness cut at $z_{850}$ and some massive galaxies owing to the
blueness cut in $V_{606} - z_{850}$.  Assuming our fiducial reddening
prescription, we find that B-dropout samples observe less than 50\%
of the stellar mass density in the universe at this epoch.  The $RJL$ 
color criterion defined by~\citet{daddi2004} does not exclude any of 
our simulated galaxies for redshifts below 4.7.

\end{itemize}

\noindent For the physical properties of simulated high-redshift galaxies, we find:

\begin{itemize}

\item Stellar mass and star formation rate are tightly correlated
(equation~\ref{eqn:sfrmass}), leading to the fundamental prediction that
GOODS B-dropouts are the most massive galaxies to have formed by $z\sim4$.

\item The B-dropout stellar masses fall predominantly in the range
$\lgmstar =$ 9--10 with a significant tail of massive galaxies out to
$\lgmstar \approx 11$, in good agreement with recent observational
assessments and somewhat more massive than the semi-analytic sample
of~\citet{Idzi1}.

\item B-dropouts are metal-enriched and obey a tight stellar mass-metallicity
relation.  The slope of this relation is in excellent agreement with the
low-redshift relation of~\citet{Tremonti2} and the normalization is
such that B-dropouts of a given stellar mass are roughly 0.6 dex less enriched 
than a typical low-redshift galaxy of the same mass. 

\item The median ages of the star particles in the B-dropouts range from
200--600 Myr, consistent with observational constraints.  However, stellar
age is not tightly correlated with any other physical or photometric parameter.

\item There is a weak correlation between stellar mass and environment
in the sense that the most massive galaxies are found in the densest
environments; more careful modeling is required to determine if this
is observable in the GOODS data.

\item A significant population is forming stars at rates in excess
of 100~$\smyr$, including several exceeding 1000~$\smyr$.  These high
star formation rates are not dramatic excursions over a slower quiescent
rate, but rather are simply the rates that are appropriate for their large
stellar masses.  The most rapid star formers have properties intriguingly
similar to SCUBA galaxies, though a firm assessment of this hypothesis
will require further work.

\item Rest-frame UV and rest-frame optical fluxes are both correlated
with star formation rate; however, owing to the tight correlation between
stellar mass and star formation rate together with the scattering effect
of dust reddening on rest-frame UV flux, rest-optical fluxes are more
tightly correlated with star formation rate than rest-UV fluxes, even
if dust reddening can be accurately removed.

\item More massive galaxies are redder.  This trend is present even
in the absence of dust reddening, although dust enhances it since more
massive galaxies are dustier in our reddening prescription.  This results
in rest-frame optical flux being correlated with color in the sense that
brighter galaxies are redder, consistent with recent observations.

\item Our simulations do not produce a bimodal distribution of galaxy
colors as seen in the red sequence at low redshift.  We have predicted
where the division might occur if the trend noted by~\citet{Bell1} at
$z < 1$ persists out to $z\sim4$ and find that many dusty star-forming
galaxies may be red enough to satisfy a red sequence color cut, but this
sequence would not be tight.

\item The distribution of galaxies on a UV-to-optical color-color plot
may be used to constrain the typical amount and form of reddening suffered
by B-dropouts although it probably cannot be used on individual galaxies.

\end{itemize}

In summary, we find that current simulations of galaxy formation broadly
pass the tests presented by available observations at $z\sim 4$, and
provide valuable insights into the nature of high redshift galaxies.
We look forward to engaging in more detailed comparisons with observations
to test galaxy formation models in more detail.

\acknowledgments
The authors thank Mark Dickinson, Rodger Thompson, Neal Katz, and David
Weinberg for highly useful discussions.  We thank Harry Ferguson and Mauro
Giavalisco for assistance with the GOODS data and for providing us with the
results from their papers.  The relation between metallicity and reddening in 
the SDSS galaxy sample was kindly provided by Christy Tremonti, whom the authors 
would also like to thank for additional advice and encouragement.   A solar 
spectrum used in computing the mass-to-light ratio was kindly provided by 
Kate Su.  This work was supported by a University of Arizona College of Science
Fellowship, an NSF Graduate Research Fellowship,
NSF grants ACI 96-19019, AST 00-71019, AST
02-06299, and AST 03-07690, and NASA ATP grants NAG5-12140,
NAG5-13292, and NAG5-13381.

%{\it Facilities:} \facility{HST (ACS)}

\end{document}